\documentclass[final]{alt2026} 

\usepackage{todonotes}
\usepackage{thm-restate}
\usepackage{enumerate}
\usepackage{enumitem}

\DeclareMathOperator{\indicate}{1 \kern -4pt 1}

\DeclareMathOperator*{\V}{\ell}

\DeclareMathOperator*{\diam}{diam}
\DeclareMathOperator*{\gdiam}{\Delta}
\DeclareMathOperator*{\core}{C}

\DeclareMathOperator*{\PP}{\mathtt{PP}}

\newcommand{\Gn}{\mathcal{G}_n} 
\newcommand{\Gnk}{\mathcal{G}_{n}^{k\text{-reg}}} 
\newcommand{\GnErdos}{\mathcal{G}(n,\frac{1}{2})} 

\newcommand{\X}{\mathcal{X}} 
\newcommand{\N}{\mathcal{N}} 
\newcommand{\M}{\mathcal{M}} 

\DeclareMathOperator*{\E}{\mathbb{E}}

\DeclareMathOperator*{\argmin}{argmin}

\DeclareMathOperator*{\Unif}{unif}

\newtheorem{observation}[theorem]{Observation}

\title[Compressibility Barriers to Data Visualization]{Compressibility Barriers to Neighborhood-Preserving \\
Data Visualization}
\usepackage{times}

 \altauthor{\Name{Szymon Snoeck} \Email{sgs2179@columbia.edu}\\
  \Name{Noah Bergam} \Email{noahbergam@cs.columbia.edu }\\
  \Name{Nakul Verma} \Email{verma@cs.columbia.edu}\\
  \addr Columbia University}

\begin{document}

\maketitle

\begin{abstract}%

To what extent is it possible to visualize high-dimensional data in two- or three-dimensional plots? We reframe this question in terms of embedding $n$-vertex graphs (representing the neighborhood structure of the input points) into metric spaces of low doubling dimension $d$ in such a way that keeps neighbors close and non-neighbors far. This notion of neighbor preservation can be understood as a considerably weaker embedding constraint than near-isometry, yet it is similarly as demanding in terms of how the minimum required dimension scales with the number of points. We show that for an overwhelming fraction of graphs, $d = \Theta(\log n)$ is both necessary and sufficient for neighbor preservation. Even sparse regular graphs, which represent more restricted neighborhood connectivity structures, typically require $d= \Omega(\log n / \log\log n)$. The landscape changes dramatically when embedding into normed spaces: general graphs become exponentially harder to embed, requiring $d=\Omega(n)$, while sparse regular graphs continue to admit $d = O(\log n)$. Finally, we study the implications of these results for visualizing data with intrinsic cluster structure. We show that graphs produced from a planted partition model with $k$ clusters on $n$ points typically require $d=\Omega(\log n)$, even when the cluster structure is salient. These results challenge the aspiration that constant-dimensional visualizations can faithfully preserve neighborhood structure.
\end{abstract}


\section{Introduction}

Visualizing a 10,000-point, 1,000-dimensional dataset in a two-dimensional plot is a bold pursuit. It is also a common practice across natural and social scientific research literature, from biology to economics to physics, where colorful UMAP and t-SNE plots have become a standard feature of data analysis \citep{dimitriadis2018t, kobak2019art, han2021resource}. In light of the well-known impossibility of low-distortion metric embeddings in constant dimensions (see e.g.\ Chapter 15 of \citet{matousek2013lectures}), the putative theoretical justification of these extreme dimension reduction procedures is that they need not preserve all minutiae 
of the input data. Instead, the argument goes, highlighting only the most basic structures, like local neighborhoods of points
, is enough for most exploratory data analysis purposes. We study the conditions under which such structure-preserving embeddings are possible. 


We demonstrate that in many scenarios it is impossible to embed a dataset
in \emph{any} metric space (let alone a Euclidean space) of constant dimension while preserving the neighborhood structure. 

Our analysis begins by re-framing low-dimensional data visualization in terms of embedding the neighborhood graph of an input into a metric space. Let $V$ be a size-\(n\) set representing our data points of interest. Let $\Gn(V)$ or simply $\Gn$ for short denote the set of all $n$-vertex unweighted, undirected, simple graphs on \(V\). Let $(\X, \rho)$ be a target metric space of interest 
with doubling dimension $d=\dim(\X)$ (see Section \ref{sec:prelims} for a definition). We think of a map $f:\Gn \rightarrow \X^n$ as a \textit{data visualization algorithm} and \(f(G)\) as an \textit{embedding} of a specific graph \(G= (V,E) \in \Gn\) into $\X$ via the algorithm \(f\).
Out of convenience, 
we write $f_G(v)$ to denote the point in $f(G)$ to which the vertex $v$ of $G$ is being mapped. If there is an edge between \(u\) and \(v\) in \(G\), we write \(u\sim_G v\), or simply \(u\sim v\) when clear from context 
(symmetrically, we write \(u\not\sim v\) if there is no direct edge). 

Within this framework, we study data visualization algorithms that are faithful to the neighborhood structure of input graphs, in the sense of keeping neighbors close and non-neighbors far. We define this desideratum below, and then study the minimum doubling dimension of the output metric space, \(\X\), necessary to accommodate it. 




\begin{definition}\label{def:preserve}
    Fix $\alpha \geq 0$. A data visualization algorithm \(f: \Gn \to \mathcal{X}^n\) is said to \(\alpha\)-preserve \(G\in \Gn\) if there exists $r=r_G > 0$ (the neighborhood threshold) such that for all distinct $u,v \in V$,
    \begin{enumerate}
        \item[(1)] $ u \sim v \implies \rho(f_G(u), f_G(v)) < r$, \hfill (neighbor proximity)
        \item[(2)] $ u \nsim v \implies \rho(f_G(u), f_G(v)) \geq \alpha\cdot r $.  \hfill (non-neighbor separation)
    \end{enumerate}
\end{definition}

\begin{figure}
    \centering
    \includegraphics[width=0.4\linewidth]{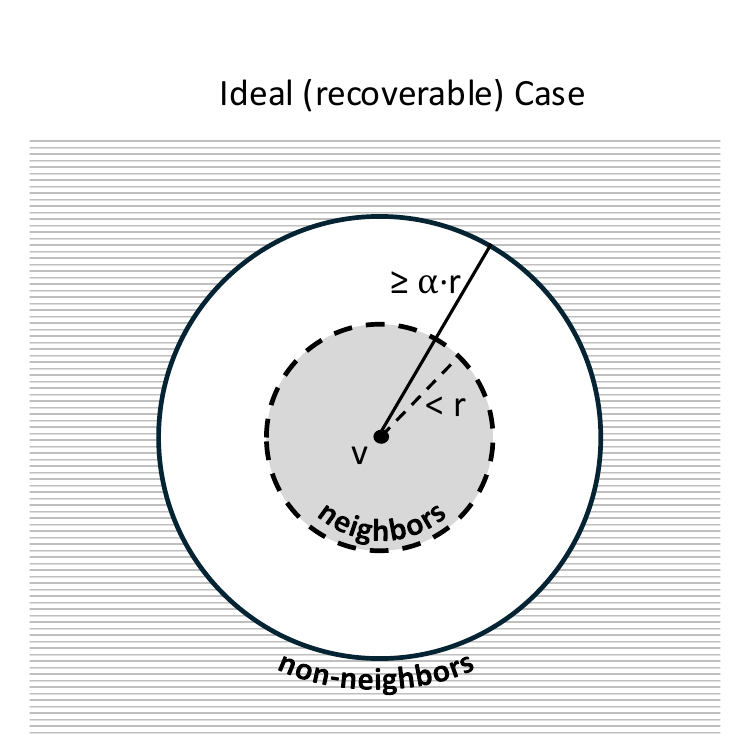}
    \includegraphics[width=0.4\linewidth]{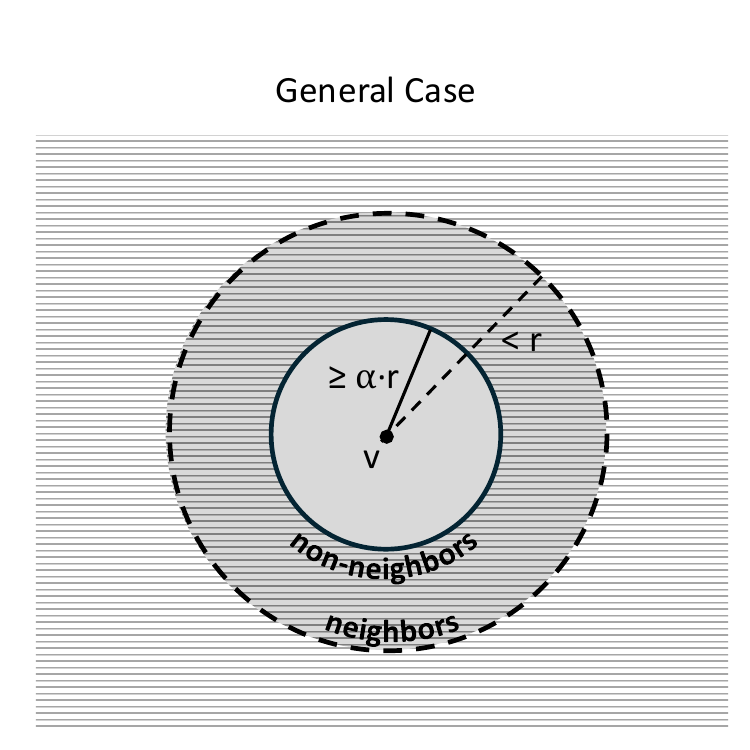}
    \caption{
    An $f$-embedding of a vertex \(v \in V\). Left: non-neighbor and neighbor relations of \(v\) are recoverable by thresholding at \(r\) \((\alpha \geq 1)\). Right: the more relaxed case of preservation where overlap is allowed \((\alpha < 1)\).
    }
    \label{fig:neighbor_overlap}
\end{figure}

Note that when $\alpha \geq 1$, one can recover the input graph \(G\) from the embedding \(f(G)\) by simply drawing an edge between any two vertices embedded within a distance of \(r\).  We call this special case of preservability \textit{recoverability}, and we notice that, depending on the setting, this distinction can have a sharp impact on the difficulty of preservability. 
Furthermore, preservability (including the special case of recoverability) stands out from other notions of structure preserving embeddings, such as ordinal embeddings, low-distortion embeddings, etc., due to its strictly local nature; it only requires the neighbor structure to be preserved, while non-neighbor structure can be distorted arbitrarily. 
Interestingly, while it seems like the locality of \(\alpha\)-preservation should make it easier to satisfy from a compressibility perspective, 
we find that \(\alpha\)-preserving a typical graph even in general metrics is as hard as near-isometric embedding of points in \(\ell_2\) (typically considered very rigid for embedding purposes) in that both require \(\Omega(\log n)\) dimensions \citep{larsen2017optimality}. 



We are interested in characterizing the minimum dimension necessary to \(\alpha\)-preserve a graph
. This is formalized in what we call the \(\alpha\)-preservation dimension. 

\begin{definition}[\(\alpha\)-preservation dimension]
    Fix $\alpha \geq 0$ and $n \in \mathbb{N}$. Let 
    $\mathbb{X}$ be a collection of metric spaces of interest. The \(\alpha\)-preservation dimension of \(G\) in \(\mathbb{X}\) is given by\footnote{In the degenerate case when no such \(f\) exists, \(\dim_\alpha(G,\mathbb{X})\) is undefined. 
    }
\begin{equation}\label{eq:alpha_pres}
      \dim_{\alpha}(G, \mathbb{X}) := \min \{ \dim(\X) \ | \ \X \in \mathbb{X} \text{ and there exists $f:\Gn \to \X^n$ which $\alpha$-preserves $G$}\}.  
    \end{equation}
    In other words, 
    it is the smallest \(d\in \mathbb{N}\) such that there exists a metric space $\X\in \mathbb{X}$ of doubling dimension \(d\) that $\alpha$-preserves $G$. 
    If $\mathbb{X}$ is the collection of all possible metric spaces, we shorten the above to $\dim_{\alpha}(G)$. 
\end{definition}

Our notion of preservation dimension can be understood as a natural generalization of a well-studied graph invariant known as \emph{sphericity} \citep{MAEHARA198455}. The sphericity of a graph \(G\) is the minimum dimension \(d\) such that the vertices of \(G\) can be distinctly embedded into a \(d\)-dimensional Euclidean space such that the embedded vertices are distance at most \(1\) if and only if they are edge-connected. Preservation dimension relaxes this notion by 
(1) parametrizing the separation between neighbors versus non-neighbors via \(\alpha\) (see Figure \ref{fig:neighbor_overlap}) 
and (2) allowing for embeddings into general metric spaces. In developing this generalization of sphericity, we obtain a more fine-grained understanding of structure-preserving metric embeddings of graphs.
\\





Our main results are as follows. 
\begin{itemize}
    \item \textbf{Preservation in General Metrics}. Though certain kinds of graphs are easily \(\alpha\)-preserved in constant dimensions (e.g.\ cliques, cycles, paths, etc.), we show that these ``easy'' graphs comprise a vanishing fraction of all graphs: an overwhelming fraction of \(G\in \Gn\) require \(\dim_\alpha(G) = \Omega
    \big(\log(n) / \log(\frac{8}{\alpha})\big)\), see Theorem \ref{thm:NeighborPreservingMain_1}(i). Even if we consider only constant-degree regular graphs, a natural model for neighborhood connectivity, the situation is similar: an overwhelming fraction of such graphs require $\Omega\big(\log(n)/(\log (\frac{\log (n)}{\alpha}))\big)$ dimensions, see Theorem \ref{thm:NeighborPreservingMain_1}(ii). We conclude with a full characterization of \(\alpha\)-preservation for \textit{all} constant-diameter graphs, see Corollary \ref{cor:metric_ub_typical}, which highlights a key difference between the cases of recoverability and non-recoverability. 
    \item \textbf{Preservation in Normed Spaces}. When we insist on embedding graphs into normed spaces, 
    the neighborhood recoverability landscape changes dramatically: an overwhelming fraction of \(G\in \Gn\) require ${\dim_{(\alpha >1)}(G) = \Omega(n/\log(\frac{8}{\alpha-1}))}$ in \emph{any} normed space, see Theorem \ref{thm:polyMain}. For Euclidean spaces, we can improve this to ${\dim_{(\alpha =1)}(G, \ell_2) = \Omega(n)}$, 
    and as \(\alpha\) exceeds \(1\), a phase-change phenomenon occurs: below a certain threshold \(\alpha\)-preservation can be achieved in dimension depending on the graph's spectral properties, and beyond this threshold \(\alpha\)-preservation may not be possible, see Proposition \ref{prop:chiGBasedUpperBound}. Meanwhile, \(k\)-regular graphs do not suffer from a \(\Omega(n)\) recoverability barrier in normed spaces; \(O(k^2 \log n)\) dimensions suffice even in \(\ell_2\), see Proposition \ref{prop:kregularUpperBound}.

    \item \textbf{Preservation of Clustered Data}. We study the preservation dimension of graphs generated from a planted partition model. 
    We find that, with high probability, \(\alpha\)-preservation requires \(\Omega\Big(\frac{ (1-\xi)\log(n) + \xi\log(k)  }{\log (8/\alpha)}\Big)\) dimensions in general metrics, where \(\xi\) is a suitable measure of the cluster salience, see Theorem \ref{thm:planted_partition}. Furthermore, if we insist on \(\alpha > 1\), we are hit with a lower bound of \(\Omega(\log(n)/ \log(\frac{1}{\alpha - 1}))\) 
    regardless of cluster salience. 
\end{itemize}

These results present a formidable barrier for existing data visualization algorithms, which typically embed input points in two- or three-dimensional Euclidean space. Per the lower bounds presented in this paper, such visualizations are doomed to misrepresent a portion of the neighborhood structure. This should be of great concern to practitioners who use these algorithms for data analysis and hypothesis generation with the expectation that they reliably reveal the neighborhood structure of high dimensional datasets. 


\section{Related Work}
Our analysis of \(\alpha\)-preservation brings insights and techniques from the graph embedding literature to bear on the problem of data visualization.

\subsection{Metric Embeddings of Graphs}


Representing graphs faithfully in metric spaces is of intense interest in computer science. There are different techniques and limitations for this endeavor depending on what metric one embeds into and what structure the embedding is supposed to preserve. For instance, if one seeks to embed a graph into Euclidean space in such a way that reflects its cluster structure, then spectral clustering is a standard choice \citep{von2007tutorial}, and it provably preserves sufficiently prominent clusters in the input \citep{ng2001spectral}. If, on the other hand, one seeks a low-distortion embedding (see Definition \ref{def:distortion}) of some metric induced by a graph (such as the shortest path metric), one can use the famous result by \citet{bourgain} that guarantees an \(O(\log n)\)-distortion embedding into $O(\log n)$-dimensional Euclidean space. If one insists on arbitrarily good average distortion-\(D\) embeddings  into normed spaces, \citet{naor2021average} showed that \(n^{\Omega(1/D)}\) dimensions is necessary in general, with constant-degree expanders providing the hard instances. The situation improves if one is willing to assume some intrinsic structure; indeed, if the input metric (derived from the graph) has doubling dimension $d$, then an $O(d)$-dimensional Euclidean embedding exists with $\mathrm{polylog}(n)$-distortion \citep{AbrahamIntrinsicDimension} and an $O(d\log \log n)$-dimensional Euclidean embedding exists with $o(\log n)$ distortion \citep{chan2010ultra}.



One could also seek out an embedding which preserves graph \emph{neighborhoods}, in the sense that edge-connected vertices are mapped as neighbors and non-edge-connected vertices are mapped as non-neighbors (for some suitable sense of neighborhood). One of the first studies in the direction was by \citet{erdos1965dimension}, who define the dimension of a graph as the minimum \(d\in\mathbb{N}\) such that there exists an injection \(f:V\rightarrow\mathbb{R}^d\) for which \(u\sim v \implies \|f(u)-f(v)\|_2 = 1\) for all \(u,v\in V\). In this setting, the distances between non-edge connected vertices is irrelevant.
\citet{MAEHARA198455} later introduced a threshold-based notion of graph dimension known as sphericity, where the condition on the embedding becomes 
\(u\sim v \iff \|f(u)- f(v)\|_2 < 1\). 
\citet{reiterman1989embeddings} showed somewhat strikingly that for \(n\geq 37\), all but \((1-1/n)\)-fraction of graphs have sphericity \(\geq n/15\). Further results have lower-bounded sphericity in terms of spectral properties of the graph adjacency matrix \citep{bilu2004monotonemapssphericitybounded}. More recently, \citet{bhattacharjee2020relations} developed multiple notions of dimension for \emph{directed} graphs, motivated by the recent surge in the interest of sequential data (e.g.\ natural language). They relate these notions of embeddability to fundamental graph properties like cyclicity and eigenspectra.  


\citet{indyk2007nearest} study a notion of local structure-preserving embedding that is motivated by approximate nearest neighbor search. They show that for input metrics with constant aspect ratio (i.e.\ the diameter over the smallest interpoint distance) or constant doubling dimension, one can achieve efficient \((1+\epsilon)\)-approximate 
nearest neighbor Euclidean embedding in $\Omega(1/\epsilon^2)$ dimensions\footnote{One should note that an analogous statement for \((1\pm \epsilon)\)-isometry for points with constant doubling dimension is known not to hold \citep{alon2003problems}.}. 
In a similar vein, \citet{bartal2011dimensionality} develop a local Johnson-Lindenstrauss-type result which embeds into \(\Omega(\log k /\epsilon^2)\)-dimensional Euclidean space and promise low distortion between any input point and its \(k\)-nearest-neighbors. 





\nocite{JMLR:v24:21-0841}

\subsection{Data Visualization and Other Applications}

Data visualization is a type of extreme dimension reduction that is focused on producing two- or three-dimensional
outputs which 
reveal cluster or local neighborhood structure.
Standard (linear) dimension reduction methods like classical multi-dimensional scaling (MDS) and random projections are generally not well-suited for this purpose: when forced into ultra-low dimensions
, these embeddings tend to destroy salient structures and display ``artifacts'' unrelated to the intrinsic structure of the data \citep{dasgupta2006concentration, diaconis2008horseshoes}.
A similar phenomenon can be said of many popular manifold learning methods like Locally Linear Embedding and Laplacian Eigenmaps 
\citep{chen2021selectingindependentcoordinatesmanifolds, 
goldberg2008manifold, venna2010information}.

On the other hand, t-SNE \citep{van2008visualizing}, UMAP \citep{mcinnes2018umap}, and related ``force-based'' embedding methods \citep{jacomy2014forceatlas2, tang2016visualizing, amid2019trimap}, 
have gained widespread popularity across the general scientific literature for their seemingly remarkable ability to visualize salient structure in high-dimensional data. 
\citet{shaham2017stochastic}, \cite{linderman2019clustering}, and \citet{arora2018analysis} were the first works showing that, for sufficiently well-clustered inputs, t-SNE does indeed output the desired cluster visualization. These results corroborate t-SNE's apparent ability to tease out global cluster structure. What about local neighborhood structure? \citet{im2018stochastic} (building on the precision-recall framework of \citet{venna2010information}) and \citet{chari2023specious} provide some practical evidence that t-SNE and UMAP are less attuned to faithfully revealing neighborhoods. 

If one seeks to embed \textit{labelled} data (for downstream prediction as well as visualization), large-margin nearest neighbors is a canonical linear technique from the Mahalonobis metric learning literature \citep{weinberger2009distance}. This method aims to alter the original representation of the data such that the nearest neighbor of any point have the same label, while differently-labelled points are separated with a  ``margin''. 
 This margin or gap is akin to our notion of separability between neighbors and non-neighbors for \((\alpha > 1)\)-preservation (see Figure \ref{fig:neighbor_overlap}, left). Similar nonlinear techniques often used in practice include contrastive learning \citep{contrastivelearning} and Siamese networks \citep{siamesenetworks}.

A fundamental question in the backdrop of these studies is: what is the minimum embedding dimension necessary to preserve local neighborhoods? Our work addresses this unifying question in a general setting, demonstrating when and how the embedding dimension scales with key properties of the input data (e.g.\ number of data points, connectivity and neighborhood structure, etc.).  

\section{Preliminaries}\label{sec:prelims}
 For $(\X, \rho)$ a metric space, let $B_r(x) \subseteq \X$ be an open ball of radius $r$ centered at $x \in \mathcal{X}$. The \textbf{doubling dimension} of $\X$, denoted $\dim(\X)$, is the smallest integer $d$ such that any open ball of radius \(r > 0\) in \(\X\) can be covered by at most \(2^d\) open balls of radius \(r/2\). 
  Let $\M(S, \epsilon)$ (the packing number)
  denote the size of the largest packing of points into $S$ such that the distance between any two points in the packing is at least
  $\epsilon$. Likewise, let $\N(S, \epsilon)$ (the covering number) denote the size of the smallest covering of $S$ by $\epsilon$-radius balls centered in $\X$. When the context is clear, we may abbreviate $\N(B_R(x), \epsilon)$ as $\N(R, \epsilon)$. 
Observe the following known results about covering and doubling dimension.
 
 \begin{observation} \label{obs:covering_number}
 For all $0 < \epsilon < R $,\  ${\N(R, \epsilon) \leq (2^{\dim(\X)})^{\lceil \log_2(R/\epsilon) \rceil} \leq (2R/\epsilon)^{\dim(\X)}.}$
 \end{observation}

\begin{observation}\label{obs:packing_to_covering} For any \(S\subseteq \X\) and \(\epsilon > 0\), we have \(\mathcal{M}(S,2\epsilon)\leq \mathcal{N}(S,\epsilon)\leq \mathcal{M}(S,\epsilon)\).
\end{observation}

 \begin{observation}\label{obs:n_point_dd}
     Any $n$-point metric space has doubling dimension at most $\lceil \log_2(n)\rceil $.
 \end{observation}

Let \(\Gnk \subseteq \Gn\) denote the set of \(k\)-regular graphs on $n$ vertices. 
For \(G = (V,E)\in \Gn\), let \(A(G) = A \in \{0,1\}^{n\times n}\) denote the adjacency matrix, $\gdiam(G)$ denote graph diameter (\(\gdiam(G):=\infty\) if \(G\) is disconnected), \(\iota(G)\subseteq V\) denote the largest independent set, and \(\kappa(G) \subseteq V\) denote the largest clique (breaking ties arbitrarily). For $V' \subseteq V,$ let \(G|_{V'} = (V', E')\) denote the subgraph of \(G\) induced by \(V'\), where \(E' = \{(u,v)\in E: u,v\in V'\}\).

For an embedding algorithm \(f\) (as defined in Introduction), we shall use \(\diam(f(G)) := \max_{u,v}\rho(f_G(u), f_G(v))\) to denote the diameter of the \(f\)-embedding of \(G\) in \(\X\).




All the omitted proofs can be found in the Appendix.

\subsection{Elementary Observations about $\alpha$-Preservation}

Before presenting our main results, we show that \(\alpha\)-preservation is only interesting for the \(\alpha \in (0,2)\) case. This is because, aside from some very trivial cases, $(\alpha \geq 2)$-preservation is impossible. In particular:

\begin{proposition}
\label{prop:alpha_gt_2}
    Let $\alpha \geq 2$. For all $G \in \Gn$, either (i) $\dim_\alpha(G)$ is undefined (i.e.\ $\alpha$-preservation is not possible), or (ii) all connected components in $G$ are cliques and \(G\) can be \(\alpha\)-preserved in \(\mathbb{R}\).
\end{proposition}

\begin{proof}
    Suppose $G=(V,E) \in \Gn$ contains a connected component that is not a clique. We will show $G$ is not $(\alpha \geq 2)$-preservable in any metric space. Let $V'\subseteq V$ be the non-clique connected component of $G$. The diameter of $G|_{V'}$ is at least $2$. Thus there exist two vertices $u,v \in V'$ such that the shortest path between them is of length exactly $2$. Let $w$ be the ``connecting'' vertex such that $u \sim w \sim v$, and assume towards contradiction there exists an algorithm $f$ that $\alpha$-preserves $G$ in some metric space $\X$ (with neighborhood threshold of $r>0$, see Definition \ref{def:preserve}). Then the maximum distance between $u$ and $v$ in the embedding is:
    $$\rho(f_G(u), f_G(v)) \leq \rho(f_G(u),f_G(w)) + \rho(f_G(w),f_G(v)) < 2 r,$$
    which implies that $u \sim v$ since $\alpha \geq 2$ implies $u \not \sim v \implies \rho(f_G(u), f_G(v)) \geq r \cdot \alpha \geq 2r $. This contradicts the fact that the shortest path between $u$ and $v$ is of length 2. Thus, an $f$ that $\alpha$-preserves $G$ cannot exist.

    Suppose all connected components of \(G\) are cliques. Observe that any visualization algorithm $f: \Gn \to \mathbb{R}$ that maps each clique to a unique point 
    in $\mathbb{R}$ $\alpha$-preserves these graphs for all $\alpha > 0$, in particular \(\alpha \geq 2\) (simply choose $r>0$ small enough). 
\end{proof}

We also observe the following intuitive monotonicity properties of \(\alpha\)-preservation. 

\begin{proposition}\label{prop:monotonicity}For \(G\in \Gn\) and a collection of metric spaces \(\mathbb{X}\), assume \(\dim_\alpha(G,\mathbb{X})\) is well-defined. The visualization dimension satisfies the following properties:
\begin{itemize}
    \item[(i)] If \(\beta \leq \alpha\) then \(\dim_\beta(G,\mathbb{X}) \leq \dim_\alpha(G,\mathbb{X})\).
    \item[(ii)] If \(\mathbb{X} \subseteq \mathbb{Y}\) then \(\dim_\alpha(G,\mathbb{Y}) \leq \dim_\alpha(G,\mathbb{X})\).
\end{itemize}
\end{proposition}



Throughout the rest of the paper we shall assume $\alpha\in(0,2)$.

\section{Preservation in General Metric Spaces}\label{sec:individual}
Take any graph \(G\in \Gn\). Arguably the most natural realization of this graph in a metric space is induced by its shortest path distances. This immediately provides an \(\alpha\)-preservation of $G$ for \textit{all} \(\alpha\in (0,2)\) in an $n$-point metric space, yielding a doubling dimension of \(O(\log n)\) 
(see Observation \ref{obs:n_point_dd}). The key question is whether this \(\log(n)\)-scaling 
is necessary. It turns out that there exist graphs which do require this scaling. In fact, 
these hard instances comprise an overwhelming fraction of $\Gn$ and persist even if we allow for a substantial overlap between neighbors and non-neighbors ($\alpha \ll 1$).
Interestingly, the situation does not improve even for graphs with low connectivity, e.g.\ \(k\)-regular graphs.  


\subsection{Lower Bounds}




\begin{restatable}{theorem}{NeighborPreservingMain} \label{thm:NeighborPreservingMain}
    For any $\alpha \in (0,2)$, we have the following.

    \begin{enumerate}
        \item[(i)] \label{thm:NeighborPreservingMain_1}
        For all $n\geq 82$, at least $1-2^{-n/5}$ fraction of $G \in \Gn$:

        $$\dim_\alpha(G) \geq \frac{\log\left(n\right) - 2\log(2)}{2\log(8/\alpha)} = \Omega \left ( \frac{\log(n)}{\log(8/\alpha)}\right).$$
        
        

        \item[(ii)] For all even integers $n\geq 6$ and $k\geq 4$, at least $1-O(n^{-\sqrt{k}+2})$ fraction of $G\in \Gnk$:
        
        $$\dim_\alpha(G) \geq  \frac{\log(n/(k+1))}{\log\left(\frac{4}{\alpha}\left\lceil \frac{\log (n-1)}{\log \left(\frac{k}{2\sqrt{k-1}+1/2}\right)}\right\rceil\right)} = \Omega\Bigg(\frac{\log(n/k)}{\log \frac{\log n}{\log k}
        + \log(4/\alpha)} \Bigg).$$
    \end{enumerate}

\end{restatable}

This incompressibility result is realized by the prevalence of graphs with high connectivity. Take for instance the star graph on \(n\) nodes: it is a diameter-2 graph with $n-1$ edges. Intuitively, a neighborhood preservation (neighbors close, non-neighbors far) of a star graph requires packing \(\Omega(n)\) points in a \(O(1)\)-diameter ball, yielding a \(\log(n)\)-type lower bound on the dimension of the target metric space. We can apply the same intuition for constant-degree expanders (e.g.\ Ramanujan graphs \citep{hoory2006expander, huang2024ramanujan}), yielding a similar result for \(\Gnk\). 

We now make this intuition precise. A key quantity in our analysis will be the notion of a clique partition of a graph  (also known as the \emph{minimum clique cover} in the literature), which is a natural measure of neighborhood connectivity.

\begin{definition}[clique partition]
For $G\in \Gn$, define the \emph{clique partition} of \(G\), denoted $P(G)$, as the smallest-sized partition\footnote{We break ties in an arbitrary but fixed manner.} of \(V\) such that for all $S \in P(G)$, $G|_S$ is a clique. 
\end{definition}
One can relate the clique partition to fundamental graph quantities which will be helpful in our discussion later.
\begin{observation} \label{prop:PGlowerbounds}
    For all $G = (V,E)$, we have 
    $\chi(G^c) = |P(G)|\geq \max \left ( |\iota(G)|, \frac{|V|}{|\kappa(G)|} \right )$, where \(\chi(G^c)\) is the chromatic number of the complement graph of \(G\). 
\end{observation}

We can quantify the difficulty of $\alpha$-preservation dimension in terms of clique partitions of the input graph $G$.



\begin{lemma}\label{lem:preservation-hammer}
    For all $G = (V,E) \in \Gn$, and all $\alpha\in  (0,2) $,\footnote{We use the convention that the maximum of an empty set in \(\mathbb{R}\) is \(-\infty\).}
    $$\dim_\alpha(G) \geq \max_{\substack{U \subseteq V \\ |U| \geq 2}}\frac{\log |P(G|_{U})|  
    }{\log(4\gdiam(G|_{U})/\alpha)}.$$   

\end{lemma}
Before proving this, it is instructive to understand the significance of the ratio: it captures our intuition that large packings (numerator) in a small space (denominator) are difficult to embed.
Consider the aforementioned star graph, which has a \(\Omega(n)\)-sized clique partition and \(O(1)\) diameter, recovering the expected \(\log(n)\)-lower bound. 
In contrast, the cycle graph–––which can clearly be \((\alpha < 2)\)-preserved in constant-dimensional \(\ell_1\) space–––has both \(\Omega(n)\)-sized clique partition and diameter, relaxing our lower bound to accommodate this case. 

\begin{proof}
Let \(f_G: V\to\X^n\) be an \(\alpha\)-preservation of \(G\) into some target metric \((\X,\rho)\) with neighbor threshold $r>0$ (see Definition \ref{def:preserve}). Note that such a map always exists (see Observation \ref{obs:trivial_UB}).
Now for any $U \subseteq V$, let $G':=G|_{U}$ be a vertex-induced subgraph. Without loss of generality, assume $G'$ has finite diameter (otherwise the bound is trivially true). We proceed via a covering argument.

   First, we show that $\diam(f_G(U)) < r \cdot \gdiam(G')$. Of course, every pair of vertices $u,v\in U$ have a path length of at most $\gdiam(G')$. A trivial application of triangle inequality shows $\rho(f_G(u), f_G(v))$ is bounded by $r\cdot \gdiam(G')$ since every successive vertex pair in the path from $u$ to $v$ needs to be within distance $r$.

   Next, we show that $\mathcal{N}(f_G(U), \alpha \cdot r/2) \geq |P(G')|$. 
   Consider any $(\alpha \cdot r/2)$-covering \(\mathcal{C} = \{c_1,\ldots,c_m\}\) of $f_G(U)$. 
    For each $i \in [m]$, iteratively define 
    $$P_i := \Big\{u \in U\ \Big|\ f_G(u) \in B_{\alpha \cdot r/2}(c_i)\Big\} \setminus \bigcup_{k < i} P_k.$$  
    Observe that $G'|_{P_i}$ is a clique, since for all distinct \(u,v\in P_i\), $\rho(f_G(u), f_G(v)) < \alpha \cdot r$, which by definition of \(\alpha\)-preservation implies \(u\sim v\). Note that \(P':=\{P_i\}_{i\in [m]}\) constitutes a clique partition of \(U\), therefore $|P(G')| \leq |P'| \leq m$.  

Therefore, by Observation \ref{obs:covering_number},
    $$|P(G')| \leq \N\left(r \cdot \gdiam(G'), \frac{\alpha \cdot r}{2} \right) \leq \left( \frac{4 \gdiam(G')}{\alpha}\right)^{\dim(\X)}.$$
Equivalently, for \emph{any} $\alpha$-preserving \(f_G:V\to\X^n\) on a metric space $\X$ and subgraph $G':=G|_U$ induced from \emph{any} $U\subseteq V$, we have 
$\dim(\X) \geq 
{\Big(\log|P(G')| \Big)}
\Big/
{\Big(\log(4\gdiam(G') / \alpha)\Big)}$
 which yields the lemma statement.
\end{proof}





The proof of Theorem \ref{thm:NeighborPreservingMain} proceeds by analyzing the key quantities \(|P(G)|\) and \(\Delta(G)\) for typical graphs in \(\Gn\). 
We think of a typical graph as being drawn uniformly at random from \(\Gn\) (equivalently, produced by the Erdős–Rényi model with parameter \(1/2\)). 
With high probability, such a graph has constant diameter and a large clique partition. The latter follows from Observation \ref{prop:PGlowerbounds} and a refined analysis of established bounds on clique numbers of Erdős–Rényi graphs \citep{frieze2016RandomGraphs}. For the \(\Gnk\) lower bound, a similar analysis yields \(\Delta(G) = O(\log n)\) and \(|P(G)| = \Omega(n)\). 
The diameter upper bound follows from well-known expander properties of \(k\)-regular graphs \citep{BollobsDiam,chung1989diameters, friedman2008proof}. The clique partition lower bound follows from analyzing \(\kappa(G)\). Formal details 
are provided in Appendix \ref{app:proof_general_metrics}.

The clique partition \(P(G)\) is a powerful concept, which helps us establish the near-worst-case difficulty of \(\alpha\)-preserving typical graphs for \emph{all} \(\alpha\in (0,2)\). In light of Corollary \ref{cor:metric_ub_typical} and Lemma \ref{lem:diam2graphs},
 it tightly characterizes the \(\alpha\)-preservation dimension of such graphs, which constitute \((1-2^{-\Omega(n)})\)-fraction of \(\Gn\). 

For the remaining (atypical) fraction of graphs, the situation can be different.  
Consider the following graph: let \(G \in \Gn\) (with \(n\) even) consist of two disjoint cliques of size \( n/2\) and exactly $n/2$ edges connecting vertices of the two cliques $1$-to-$1$. 
A straightforward packing argument gives $\dim_\alpha(G)=\Omega(\log(n))$ for all $\alpha = 1 + \Omega(1)$, yet our key Lemma 
\ref{lem:preservation-hammer} is rendered ineffective since \(|P(G)| = 2\) and \(\gdiam(G) = 2\).

This example highlights how, as \(\alpha\) grows past unity
, the problem of \(\alpha\)-preservation can become significantly harder for certain graphs. 
The key difficulty of \((\alpha> 1)\)-preserving this graph stems from the fact that there exist a large number of vertices in close proximity that have \textit{distinct, yet overlapping} neighborhoods. To formalize this, we introduce the concept of a ``neighborhood partition,'' which we leverage in Lemma \ref{lem:HAMMER2}, improving on Lemma \ref{lem:preservation-hammer} when \(\alpha = 1 + \Omega(1)\). 

\begin{definition}[neighborhood partition] \label{def:CG}
For $G = (V,E) \in \Gn$ and $U \subseteq V$, define the \emph{neighborhood partition} of \(G|_U\), denoted $\core(G|_U)$, as the smallest-sized partition\footnote{We break ties in an arbitrary but fixed manner.} of \(U\) such that all $u$ and  $u'$ are in the same part if and only if they have identical neighborhoods with respect to $G$. That is, for all $u,u' \in S \in \core(G|_U)$, we have $N(u) = N(u')$, where $N(u) := \{ v \in V: v \sim_G u \text{ or } v=u\}$.
\end{definition}


\begin{lemma}
\label{lem:HAMMER2}
For all \(G = (V,E) \in \Gn\) and \(\alpha \in (1,2)\),
\[\dim_\alpha(G) \geq \;\max_{\substack{U\subseteq V \\ |U| \geq 2}}\;\frac{\log \big| \core(G|_U) \big|}{\log(\frac{4\gdiam(G|_U)}{\alpha - 1})}.\]

\end{lemma} 
\begin{proof}
It suffices to exhibit the lower bound for all \(U\subseteq V\) with $G|_U$ connected and $|\core(G|_U)|  \geq 2$ otherwise the bound is vacuously true. Fix any \(\alpha \in (1,2)\) and consider an \(\alpha\)-preservation \(f_G\) of \(G\) in $(\X,\rho)$. Note that for all \(u',u''\in V\) which are not in the same part of $\core(G|_U)$, there exists \(v \in V\) that is in either $N(u')$ or $N(u'')$ but not both since $N(u') \neq N(u'')$. WLOG assume $v \in N(u') \setminus N(u'')$. By definition of \(\alpha\)-preservation, this tells us:
    \[\alpha r \leq \rho(f_G(u''),f_G(v))\leq \rho(f_G(u'),f_G(v)) + \rho(f_G(u'),f_G(u'')) < \rho(f_G(u'),f_G(u'')) + r\]
    \[\implies \forall u',u'' \text{ in distinct parts of }C(G|_U) \hspace{1cm}\rho(f_G(u'),f_G(u'')) > r(\alpha - 1),\]
    where \(r\) is the neighborhood threshold of $f_G$ 
    (cf.\ Definition \ref{def:preserve}). 
    Since \(f_G(U)\) has diameter at most \(r\cdot \gdiam(G|_U)\), it contains an \(r(\alpha-1)\)-packing of a ball of diameter \(r\cdot\gdiam( G|_U) \) of size \(|C(G|_U)|\)
    . Due to maximal packing estimates (cf.\ Observation \ref{obs:packing_to_covering})
    , we have:
    $$| \core(G|_U)| \leq \mathcal{M}\left(r\cdot \gdiam(G|_U), r(\alpha - 1)\right) \leq \mathcal{N}\left(r\cdot \gdiam(G|_U), \frac{r(\alpha - 1)}{2}\right) \leq \left(\frac{4\gdiam(G|_U)}{ (\alpha - 1)}\right)^{\dim(\X)}.
    $$
Rearranging the terms yields the lower bound.
\end{proof}

\subsection{Upper Bounds}

As discussed at the beginning of the section, one trivially has an upper bound on the $\alpha$-preservation dimension of a graph $G$ in terms of the doubling dimension of the graph shortest path metric \(\rho_G\).

\begin{observation}\label{obs:trivial_UB}
   For all $G \in \Gn$ and $\alpha \in (0,2)$, $\dim_\alpha(G) \leq \dim(\mathcal{X}_{\rho_G}) \leq \lceil \log_2 n \rceil $, where $\mathcal{X}_{\rho_G}$ is the $n$-point shortest path metric derived from $G$. 
\end{observation}

Observe that 
this upper bound in terms of the shortest path metric is clearly not tight. Consider for instance the complete graph on \(n\) vertices. The doubling dimension (with respect to the shortest path metric) of this graph is \(\Theta(\log n)\), yet it can be \(\alpha\)-preserved in constant dimensions, by simply sending all points to a one-point metric space. It is natural to ask whether an \(\alpha\)-dependent bound is possible. 

\begin{restatable}{proposition}{NontrivialUBPG} 
\label{prop:nontrivial_PG_ub} 
    For all \(G\in \Gn\),
\begin{equation}
\nonumber
\dim_\alpha(G) \leq 
    \left\{
    \begin{array}{ll}
    \Big\lceil\log_2(3)\Big\lceil \frac{ \log |P(G)|}{\log \lceil 1/\alpha\rceil}\Big\rceil \Big\rceil  & \alpha\in(0,1)  \vspace{0.05in} \\ 
 \big\lceil\log_2 |P(G)| + \log_2(3)\big\rceil & \alpha=1  \vspace{0.05in} \\    
   \Big\lceil\log_2 |P(G)| + \log_2(3) \Big\lceil \max_{S\in P(G)} \frac{\log |\core(G|_S)|}{\log \lceil \frac{1}{\alpha-1}\rceil} \Big\rceil \Big\rceil
   & \alpha\in(1,2) 
    \end{array}
    \right.
.
\end{equation}
\end{restatable}
The proof follows an intuitive construction which hinges on the clique partition. For \(\alpha < 1\), each part of the clique partition can be collapsed to a point yielding a \(|P(G)|\)-sized metric space of doubling dimension \(O(\log |P(G)|)\). For \(\alpha \geq 1\), each part \(S\) is embedded as a sub-metric which necessarily contains some \(\alpha-1\) packing depending on its neighborhood partition \(C(G|_S)\). Compare this with our lower bounds: Lemma \ref{lem:preservation-hammer} for the \(\alpha \leq 1\) case and Lemma \ref{lem:HAMMER2} for the \(\alpha > 1\) case.



This enables us to give a full characterization of \(\alpha\)-preservation dimension for constant-diameter graphs, which by Lemma \ref{lem:diam2graphs} constitute an overwhelming fraction of \(\Gn\).



\begin{corollary}
\label{cor:metric_ub_typical}
For \(\alpha\in (0,2)\) and \( G \in \Gn\) such that \(\Delta(G) = O(1)\), 
\[\dim_\alpha(G) = \Theta\Big(\frac{\log|P(G)|}{\log(8/\alpha)} + \indicate[\alpha>1] \cdot \frac{\log|C(G)|}{\log(\frac{4}{\alpha-1})}\Big). \]
\end{corollary}

\section{Preservation in Normed Spaces}\label{sec:families}


The designation of metric space is incredibly general. Practitioners are usually interested in producing faithful data visualizations in more structured spaces. For instance, it is often desirable to work with data in a normed space due to the ability to add and scale the points as vectors. The significance of the norm structure comes from not only its mathematical convenience but also its direct interpretability for data visualization: we tend to think in terms of normed (especially Euclidean) space because it obeys similar principles as physical space.
This motivates us to study $\alpha$-preservation into such spaces. We find that this restriction comes at a steep cost in terms of the embedding dimension.



\subsection{Lower Bounds}



It turns out that \((\alpha\geq 1)\)-preservation (that is neighborhood \emph{recoverability}) in normed spaces is exponentially harder than it is for general metrics: an overwhelming fraction of graphs in $\Gn$ require dimension that scales \textit{linearly} in \(n\), in contrast to the 
logarithmic scaling in the case of general metrics (cf.\ Theorem \ref{thm:NeighborPreservingMain}).


\begin{restatable}{theorem}
{polyMain}
\label{thm:polyMain}
We have the following.
\begin{enumerate}[label=(\roman*)]
    \item \textbf{(General normed spaces)} Let $\mathbb{L}$ be the collection of all normed spaces.  For all $\alpha \in (1,2)$ and $n\geq 82$, we have that for at least $1-2^{-n/6}$ fraction of $G \in \Gn$:
$$
\dim_\alpha(G,\mathbb{L})  \geq \frac{n}{3\log_2(\frac{16}{\alpha - 1})} = \Omega\Bigg(\frac{n}{\log(\frac{16}{\alpha-1})}\Bigg).
$$
\item \textbf{(Euclidean spaces)} For $\alpha = 1$ 
and $n \geq 0$, we have that for at least $1-2^{-n}$ fraction of $G\in \Gn$, 
$$
\dim_{(\alpha=1)}(G,\ell_2) \geq \frac{n}{15}-\frac{1}{4} = \Omega(n).
$$
\end{enumerate}
\end{restatable}

Before discussing the proof of these dramatic lower bounds, it is important to distinguish where they do \textit{not} apply.
\begin{itemize}
\item There is no \(\Omega(n)\) lower bound for $(\alpha <1)$-preservation in normed spaces. It turns out that \emph{all} graphs can be $(\alpha<1)$-preserved in $\ell_\infty^d$ with $d = O(\log n)$ (see Proposition \ref{prop:linf_ub}).
\item For \(k\)-regular graphs, there is no \(\Omega(n)\) lower bound for \((\alpha\geq 1)\)-preservation in normed spaces. 
In particular, one can show that an $O(k^2 \log n)$-dimensional $\ell_2$ embedding exists for all such graphs (see Proposition \ref{prop:kregularUpperBound}) which nearly matches the lower bounds of Theorem \ref{thm:NeighborPreservingMain}(ii). 
\end{itemize}

The lower bound on \((\alpha > 1)\)-preservation dimension in general normed spaces proceeds via an elegant application of the volume argument. 
Specifically, the vector space structure of normed spaces enables us to superimpose the sheer plenitude of low-diameter graphs in a small region. Noting that disparate graphs (with disparate neighborhood structures) require disparate neighborhood-preserving embeddings in the space, one must require the target dimension to scale with $n$ to accommodate these embeddings. Formal details are provided in Appendix \ref{app:proof_normed_spaces}.

The highly regular structure of Euclidean space enables us to extend our lower bound to the $\alpha=1$ case in $\ell_2$. This requires a very different analysis. In particular, we leverage the fact that since the (squared) $2$-norm distances can be represented by a (quadratic) polynomial, any graph can be distinctly recognized from an $(\alpha=1)$-preserving embedding in $\ell_2$ by a series of polynomial threshold tests. By a theorem due to \citet{WarrensPoly}, we know that the polynomial threshold tests are expressive enough to distinguish the graphs only if the embedding dimension is $\Omega(n)$, yielding the desired extension.\footnote{This method of proof extends to other $p$-norms (for $p$ even), and may be of independent interest.} 
Formal details are provided in Appendix \ref{app:proof_normed_spaces}.

\subsection{Upper Bounds for Preservation in Normed Spaces}

A simple application of Fréchet's embedding \citep{frechet, matousek2013lectures} yields an isometric embedding of the shortest path metric (hence an \((\alpha \leq 2)\)-preservation) of the input graph into $\ell_\infty^{n-1}$, thereby telling us ${\dim_\alpha(\Gn, \mathbb{L}) = O(n)}$. Specifically:

\begin{observation}\label{prop:frechetEmbedding}
  For all \(\alpha\in (0,2)\) and  \(G\in \Gn\), \(\dim_\alpha(G, \ell_\infty) \leq \lceil \log_2(3)\cdot  (n-1) \rceil = O(n)\).
\end{observation}
%
This trivial result can be refined as follows. 
\begin{proposition}
\label{prop:linf_ub}
    For all $G\in \Gn$,
    $$
    \dim_\alpha(G,\ell_\infty) \leq \left \{
    \begin{array}{ll}
        \Big\lceil\log_2(3)\Big\lceil \frac{ \log |P(G)|}{\log \lceil 1/\alpha\rceil}\Big\rceil \Big\rceil =O\Big(\frac{\log |P(G)|}{ \log\lceil1/\alpha\rceil}\Big) & \alpha \in (0,1) \vspace{0.05in}\\
        \lceil \log_2(3) \cdot |C(G)| \rceil = O(|C(G)|) & \alpha \in [1,2)
    \end{array} 
    \right. .$$
\end{proposition}
\begin{proof} For any \(G=(V,E)\in \Gn\), let $P(G)=\{P_1,\ldots,P_m\}$ be the $m$ parts of the clique partition.

\emph{Case $\alpha \in (0,1).$}
By Observation \ref{obs:grid_packing}(i) (take $n=m$, $r=1$ and $\epsilon = \alpha$), $m$ points can be embedded in (open) unit ball with interpoint distances at least $\alpha$ in $\ell_\infty^d$ with $d=\big\lceil \frac{\log m}{\log\lceil 1/\alpha\rceil} \big\rceil$. Thus, the mapping where vertices belonging to each partition $P_i$ of the input graph $G$ is mapped to the $i$-th point yields an $\alpha$-preserving embedding in $\ell_\infty$, with (doubling) dimension at most $\Big\lceil\log_2(3)\Big\lceil \frac{ \log |P(G)|}{\log \lceil 1/\alpha\rceil}\Big\rceil \Big\rceil
= O\Big(\frac{\log |P(G)|}{\log \lceil 1/\alpha\rceil}\Big)$.

\emph{Case $\alpha \in [1,2).$} Let $G / C(G)$ be the (simple) graph produced by contracting all nodes that are in the same part of $C(G)$ and preserving the edge connectivity. Since any $u,v \in V$ with $N(v)=N(u)$ (see Definition \ref{def:CG}) can be embedded identically with no effect on $\alpha$-preservation, an \(\alpha\)-preservation of \(G/C(G)\) is an \(\alpha\)-preservation of \(G\). Hence we re-apply Observation \ref{prop:frechetEmbedding} to $G / C(G)$ to get the bound\footnote{Note that this embedding works for \(\alpha\in (0,2)\), but since \(|P(G)| \leq |C(G)|\) it offers no improvement.} $\lceil \log_2(3) \cdot |C(G)| \rceil$.
\end{proof}




For the $\alpha < 1$ case, this upper bound improves considerably on Fréchet-style embedding (Observation \ref{prop:frechetEmbedding}) and nearly matches the general metric space lower bound given by Lemma \ref{lem:preservation-hammer}. On the other hand, in the recoverable case, one cannot hope for more than a constant-factor improvement on this straightforward upper bound due to a result by \citet{BoxicityRobert1969} who proved that there exist $G \in \Gn$ with $\dim_{\alpha \geq 1}(G, \ell_\infty) \geq \left\lfloor \frac{2}{3}|C(G)| \right\rfloor.$

\subsubsection{Preservation in Euclidean Space ($\ell_2$)}

Theorem \ref{thm:polyMain}(ii) establishes a formidable \(\Omega(n)\) lower bound on \(\ell_2\) recoverability for most \(G\in \Gn\). A matching \(O(n)\) upper bound is trivial: if an \(\alpha\)-preservation of \(n\) points exists, it is necessarily an \(\alpha\)-preservation on the \((n-1)\)-dimensional subspace spanned by the points. 
Here we investigate more refined upper bounds depending on \(\alpha\) as well as the structure of the input graph.  

We find that \((\alpha < 1)\)-preservation dimension in \(\ell_2\) follows a familiar \(\log |P(G)|\) scaling. In contrast, the recoverable case exhibits an interesting phase shift. For \(\alpha\) up to a graph-dependent threshold, the graph is recoverable in dimension depending on its spectrum. However, past this threshold, \(\alpha\)-preservation is not always possible. In particular: 


\begin{restatable}{proposition}{chiGBasedUpperBound}
\label{prop:chiGBasedUpperBound}
For all \(G\in \Gn\), we have
    $$
    \dim_\alpha(G,\ell_2) \leq 
    \left\{
    \begin{array}{ll}
    \Big\lceil\log_2(5) \Big\lceil \frac{ 4 \log (|P(G)|+1)}{2-4\alpha^2}\Big\rceil \Big\rceil
    & \alpha\in(0,\frac{1}{\sqrt{2}})  \vspace{0.05in} \\ 

    \Big\lceil\log_2(5) \Big\lceil 12 \big( \frac{1+\alpha^2}{1-\alpha^2}\big)^2 \log |P(G)\Big\rceil \Big\rceil& \alpha\in \big( \frac{1}{\sqrt{3}},1 \big)  \vspace{0.05in} \\ 

   \Big\lceil\log_2(5) \cdot \min\big( \lceil 192\lambda_G^2 \log|C(G)| \rceil ,  \  |C(G)|-1 \big) \Big\rceil  & \alpha\in \Big[1,\frac{1}{\sqrt{1-\frac{1}{4\lambda_G}
   }} \Big)  \vspace{0.05in} 
    
    \end{array}
    \right. ,
    $$
    
where \(\lambda_G\) denotes the maximum eigenvalue of \(A(G / C(G))\). 

Moreover, for all \(n\geq  4\) there exists \(G\in \Gn\) which cannot be \(\alpha\)-preserved in \(\ell_2\) for \(\alpha > (1  - \frac{1}{\lambda_G})^{-1/2}\).
\end{restatable}



In light of Theorem \ref{thm:polyMain}(ii) it is worth noting that only a negligible fraction of \(G\in \Gn\) have \(\lambda_G = o(n)\). Since \(\lambda_G\) is upper-bounded by the maximum vertex degree, Proposition \ref{prop:chiGBasedUpperBound} immediately provides us a recoverability upper bound \(k\)-regular graphs.
\begin{corollary}
\label{prop:kregularUpperBound}
    Fix any $0 < k < n$. For all \(G\in \Gnk\) and $\alpha\in(0,\sqrt{1+\frac{1}{4k}})$, we have, $
    \dim_\alpha(G,\ell_2) \leq \Big\lceil\log_2(5) \cdot\lceil 192 k^2 \log n \rceil \big \rceil. $
\end{corollary}



\section{Preservation of Clustered Data}
\label{sec:nbhd}




Most data visualization techniques seek to reveal underlying cluster structure. One may hope that the presence of a salient clustering would make neighborhood preservation more tractable. We show, for graphs sampled from a planted partition model, this is largely not the case.  

\begin{definition}\label{def:pp_model}
    Fix integers \(n\geq k \geq 1\). For a partition $\{S_1, \ldots, S_k\}$ of $V$ 
    and ${0 \leq q \leq p \leq 1}$, define the planted partition distribution, $\PP_{p,q}(S_1,\ldots,S_k)$, as the distribution over $\Gn(V)$ such that for all $i,j \in [k]$ distinct, edges within $S_i$ occur independently with probability $p$ and edges between $S_i$ and $S_j$ occur independently with probability $q$.
\end{definition}

The key parameters \(p\) and \(q\) capture intra- and inter-cluster connectivity. Naturally, the higher the gap between these two parameters, the more salient the cluster structure of the model. We provide a lower bound on the \(\alpha\)-preservation dimension with respect to these parameters.  


\begin{restatable}{theorem}
{plantedPartition}
\label{thm:planted_partition}
   Fix $ n \geq k\geq 1$ and let $\{S_1, \ldots, S_k\}$ be a partition of $V$ (of size $n$). Let $c\geq1$ be such that ${\max_{i\in[k]} |S_i| \leq \frac{cn}{k}}$, and  $0 < q \leq p \leq 1$ with $q<1$. Then for all $\alpha \in (0,2)$ with probability at least ${1-\exp(-\Omega(n^{\min(2(p+q-pq),1)}))}$ over $G \sim \PP_{p,q}(S_1,\ldots,S_k)$:
   \begin{equation*}
       \dim_\alpha(G) \geq \frac{1}{\log\left(8/\alpha\right)}\Big( (1-\xi_{p,q}) \log n + \xi_{p,q} \log(k/2c)    \Big), 
   \end{equation*}
where $\xi_{p,q} :=p-q+pq$ encodes the \emph{cluster saliency} of the planted partition model. 
\end{restatable}

The proof proceeds by estimating the clique number and the diameter of typical graphs generated from a planted partition model and applying our key Lemma \ref{lem:preservation-hammer}. Formal details are provided in Appendix \ref{app:cluster}.


We can think about this result as interpolating between two regimes:

\begin{itemize}
\item \textit{Maximally clustered}. When \(0 < q < p = 1\) (with \(q\) constant), we have \(\xi_{p,q}=1\), so the lower bound becomes $\Omega(\frac{\log(k/2c
)}{ \log(8/\alpha)})$. This essentially matches our upper bound in Proposition \ref{prop:nontrivial_PG_ub} for \(\alpha\leq 1\), since \(|P(G)| = k\).
    \item \textit{Maximally unclustered.} When $p=q=\frac{1}{2}$, the lower bound  becomes \(
    \Omega(\frac{\log(n)}{ \log(8/\alpha)})\), which matches the upper bound in  Proposition \ref{prop:nontrivial_PG_ub}. 
\end{itemize}


For \(\alpha > 1\), we have a stronger lower bound: even when \(p=1\) (and hence clusters are fully connected), the \(\alpha\)-preservation dimension is \(\Omega(\log n)\) with overwhelming probability. 

\begin{proposition}\label{prop:clusters_alpha_geq_1}
   Fix any $\alpha > 1$. Pick $p=1$ and $0<q<1$. Then if \(\max_{i\in [k]}|S_i| \leq c n/k\), \(G\sim \PP_{p,q}(S_1,...,S_k)\) with probability at least $1 - n^2\left(\max(q,1-q)^{2n(1-c/k)}+e^{-q^2(n-1)}\right)$, we have $\dim_{\alpha}(G) \geq \frac{\log(n)}{\log(\frac{8}{\alpha - 1})}.$
\end{proposition}

\section{Discussion}\label{sec:discussion}



Given the pervasive use of data visualization techniques like t-SNE and UMAP which ``just seem to work'' in practice, it is tempting to believe that data visualization is essentially a solved problem. Our analysis of \(\alpha\)-preservation reveals some evidence to the contrary: in many situations, including when data has extremely pronounced cluster structure, visualizing neighbor and non-neighbor relationships fundamentally requires high dimensions.


One may wonder how the incompressibility results presented in this work square with more positive findings regarding the possibility of constant-dimensional data visualizations. \citet{sarkar2011low}, for instance, showed that one can embed trees near-isometrically in a two-dimensional hyperbolic space.
 This may seem to contradict Lemma \ref{lem:preservation-hammer}, which asserts that trees have \(\Omega(\log n)\) preservation dimension in general. Note, however, that a constant-dimensional hyperbolic space has infinite \emph{doubling} dimension. Other works, such as that of \citet{arora2018analysis}, established that two-dimensional t-SNE plots faithfully depict the overall cluster structure of well-clustered inputs. While this may appear to be at odds with Theorem \ref{thm:planted_partition}, our result are concerned with the preservation of \textit{local} neighborhoods as opposed to \textit{global} cluster structure. 

One can think of our notion of neighborhood preservation as a natural relaxation of the more well-studied notion of low-distortion embedding. Indeed, a \((1 + \epsilon)\)-distortion embedding is also a \((\frac{1}{1 + \epsilon})\)-preservation, see Observation \ref{obs:DistortionImpliesPreservationMetric}. Preservation is refreshingly lenient compared to low-distortion. For instance, \(1\)-preservation (the hardest form of \(\alpha\leq 1\) preservation) is possible for any graph in \(\ell_2\), whereas \(1\)-distortion (the hardest form of low-distortion embedding) is very much not possible: even some graph metrics of constant doubling dimension require \(\sqrt{\log n}\)-distortion in \(\ell_2\) \citep{gupta2003bounded}! 


There are many avenues 
for further inquiry. For instance, what properties of data \mbox{enable} constant-dimensional neighborhood preservations? We showed that sparse and well-clustered \mbox{inputs} typically require \(\Omega(\log n)\) dimensions to preserve. Considering the local nature of our embedding criterion, it is worth investigating its interplay with manifold-type \mbox{inputs}.
Another natural line of investigation involves approximate \(\alpha\)-preservation: given some neighborhood graph and a fixed target dimension $d$, what fraction of neighborhoods can be $\alpha$-preserved in $d$? Does this relaxation dramatically change the stringent 
lower bounds of this work? 
If so, then answering this question for real-world datasets would allow us calibrate our expectations for data visualization on a case-by-case basis. 
Finally, it is tempting to pursue an algorithmic realization of \(\alpha\)-preservation. Our study of \(\alpha\)-preservation suggests that certain efficiently computable graph statistics, including \(C(G)\) and the graph spectrum, can help to quantify \(\alpha\)-preservation dimension.

The desire to \textit{see} neighborhood relationships in data is ubiquitous. 
Our work reveals serious geometric barriers to this pursuit.  





\acks{N.B. was supported by the ONR grant N0001424SB001 and The Herbert and Florence Irving Institute for Cancer Dynamics during the writing of this paper.}

\vskip 0.2in
\bibliography{biblio}

\newpage
\appendix

\section{Useful Supporting Results}

\begin{restatable}{lemma}{FamDiamBounds} 
\label{lem:diam2graphs}
Fix any $q\in(0,1]$. Let $\mathcal{D}_G$ denote a distribution over \(\Gn(V)\) where the each edge appears independently with probability at least $q$. Then \(\mathbb{P}_{G\sim \mathcal{D}_G }\Big[\Delta(G) \leq 2 \Big] \geq 1-n^2e^{-q^2(n-1)} \geq 1 - 2^{-\Omega(n)}\).
\end{restatable}
\begin{proof}\textbf{of Lemma \ref{lem:diam2graphs}} The probability that any distinct $u,v\in V$ have distance strictly greater than $2$ is the probability that they do not share an edge and have no common neighbor, which  (by independence of edges) is 
$\leq (1-q) \cdot (1-q^2)^{n-2}$. By union bounding over all pairs of vertices, 
$\mathbb{P}[\gdiam(G)> 2] \leq \binom{n}{2}(1-q) \cdot (1-q^2)^{n-2}
\leq n^2 e^{-q^2(n-1)}.$
\end{proof}

\begin{corollary}
\label{cor:gn_diam2}
Let $S\subset \Gn$ be the subset of all graphs with diameter at most $2$, then $|S|\geq (1-n^2e^{-(n-1)/4})|\Gn|$. In particular, if $n\geq 82$, $|S| \geq (1-2^{-n/5})|\Gn| = (1-2^{-\Omega(n)})|\Gn|$.
\end{corollary}
\begin{proof}\textbf{of Corollary \ref{cor:gn_diam2}}
Recall that the uniform distribution over $\Gn$ is equivalent to the Erdős–Rényi model over $n$ nodes, $\GnErdos$, where each edge appears independently with probability $\frac{1}{2}$.

Invoking Lemma \ref{lem:diam2graphs} (with $q=1/2$) gives us
that with probability at least $1-n^2e^{-(n-1)/4}$ over a graph $G$ drawn uniformly from $\Gn$, we have that $\gdiam(G) \leq 2$.
\end{proof}

%

\begin{lemma} \label{lem:kregdiam}
For \(k\geq 4\) and \(n\geq 6\) even integers, there exists a universal constant \(c>0\) such that

    $$\mathbb{P}_{G\sim \Unif(\Gnk)}\left[\Delta(G) \leq \left\lceil \frac{\log (n-1)}{\log \left(\frac{k}{2\sqrt{k-1}+1/2}\right)}\right\rceil \right] \geq 1-cn^{-\sqrt{k}+2}.$$

\end{lemma} 
\begin{proof}\textbf{of Lemma \ref{lem:kregdiam}}
    Let $\lambda_2(G)$ be the second largest eigenvalue of $G$'s adjacency matrix. By \citet{chung1989diameters} we have \({\gdiam(G) \leq \lceil \frac{\log (n-1)}{\log (k/\lambda_2(G))}\rceil }\).  By \citet{friedman2008proof}, for \(k\geq 4\) and \(n\geq 6\) even integers, there exists a constant \(c > 0\) such that with probability $1-cn^{-\lceil\sqrt{k-1}\rceil + 1} \geq  1-cn^{-\sqrt{k}+2}$, a random regular graph has \(\lambda_2(G) \leq 2\sqrt{k-1}+1/2\). Combining these results, we have with high probability that $\gdiam(G) \leq \left\lceil \frac{\log (n-1)}{\log \left(\frac{k}{2\sqrt{k-1}+1/2}\right)}\right\rceil.$
\end{proof}


    

%


\begin{lemma}[Johnson-Lindenstrauss; see \citet{dasgupta2003elementary}]\label{lem:JL} For any $0<\epsilon\leq\frac{1}{2}$ and any integer $n$, let $k$ be a positive integer such that $k\geq \frac{12 \log n}{\epsilon^2}$.
Then for any set $V$ of $n$ points in $\mathbb{R}^d$, there is a map $f : \mathbb{R}^d \rightarrow \mathbb{R}^k$ such that for all $u,v \in V$,
$$(1-\epsilon)\|u-v\|^2_2 \leq \|f(u)-f(v)\|^2_2 \leq(1+\epsilon)\|u-v\|^2_2.$$
\end{lemma}

\begin{observation}\label{obs:grid_packing} 
For any $0<\epsilon < r$, we have the following.
\begin{enumerate}[label=(\roman*)]
    \item One can always  \(\epsilon\)-pack \(n\) points in a diameter \(r\) (open) ball in $\ell_\infty^d$ with $d=\left\lceil\frac{\log n}{\log \lceil r/\epsilon\rceil}\right\rceil$.
    \item One can always  \(\epsilon\)-pack \(n\) points in a diameter \(r\) (open) ball in $\ell_2^d$ with  $d =\left\lceil\frac{4\log (n+1)}{2- (2\epsilon/r)^2}\right\rceil$.
\end{enumerate} 
\end{observation}
\begin{proof}\textbf{of Observation \ref{obs:grid_packing}(i)}
    Consider an $\epsilon$-resolution grid of an $r$ diameter open ball in $\ell_\infty^d$ for $d = \lceil\frac{\log n}{\log \left\lceil r/\epsilon\right\rceil}\rceil$. Specifically, consider the grid points 
    \(\big\{{\epsilon \cdot j: j\in\mathbb{Z}, 0 \leq j < r/\epsilon \big\}^{d}}\). Note that by construction, diameter is strictly less than $r$ and the number of grid points is at least $n$. Any subset of $n$ points from this grid is an $\epsilon$-packing (that is, for any $p,p'$ distinct from the grid, $\|p-p'\|_\infty\geq \epsilon$).
\end{proof}

\begin{proof}\textbf{of Observation \ref{obs:grid_packing}(ii)} We'll show that $n$ points can be $\gamma$-packed in a radius $1$ (i.e.\ diameter $2$) open ball of in $\ell_2^d$ for $d$ at most $\lceil \frac{4\log (n+1)}{ 2-\gamma^2}\rceil$, this immediately yields the desired result.

Fix any $\delta>0$ and consider the open ball of radius $1+\delta$ in $\ell_2^d$. It suffices to show that there exists a $\gamma$-packing of at least $n$ points on the unit sphere $S^{d-1}$ for $d$ sufficiently large. Let $\sigma$ be the uniform probability measure over $S^{d-1}$ and $\{p_1, \dots, p_m\}$ be a maximally sized $\gamma$-packing of $S^{d-1}$. For any $p \in S^{d-1}$, define ${C_\gamma(p) := \{q \in S^{d-1} : \| p-q\|_2 < \gamma\}}$ as the $\gamma$-spherical cap of $S^{d-1}$ at $p$. Observe that ${C_\gamma(p) = \left\{q \in S^{d-1} : p\cdot q > 1 - \frac{\gamma^2}{2}\right\}}$ and that $\bigcup_{i \in [m]} C_\gamma(p_i) = S^{d-1}$. Hence:
$$m\cdot \sigma(C_\gamma(p_1)) \geq \sigma\left(\bigcup_{i \in [m]} C_\gamma(p_i) \right)= \sigma(S^{d-1}) = 1.$$
Noting that the volume of a $\gamma$-spherical cap ${\sigma(C_\gamma(p_1)) \leq \exp \left(-\frac{d}{4}(2-\gamma^2)\right)}$ \citep{sphericalcaps}, and by letting $\delta \rightarrow 0$, we have that having $d =
\left\lceil\frac{4\log (n+1)}{2- \gamma^2}\right\rceil$ 
is sufficient to $\gamma$-pack $n$ points in an open unit ball in $\ell_2^d$. 
\end{proof}

\subsection*{Low-Distortion Implies Good Preservation}
In order to directly compare $\alpha$-preservation and $\frac{1}{\alpha}$-distortion embeddings, we broaden the definition of $\alpha$-preservation to apply to metric spaces:

\begin{definition}
    For a metric space \((\mathcal{Z},\sigma)\), let \(G_R(\mathcal{Z}) = (\mathcal{Z}, E)\) denote the graph where, for any $u,v \in \mathcal{Z},$ \((u,v) \in E \iff {\sigma(u,v) < R}\). We say a map \(f:(\mathcal{Z},\sigma)\to (\X,\rho)\) is an \((\alpha,R)\)-preservation of \((\mathcal{Z},\rho)\) for \(\alpha\in (0,2)\) and \(R>0\) if its output is an \(\alpha\)-preservation of \(G_R(\mathcal{Z})\).
\end{definition}

Note that $(\alpha, 2)$-preserving $G \in \Gn$ with respect to the shortest path metric coincides precisely with the regular definition of $\alpha$-preserving $G$. Consider the classical notion of structure retaining embedding:

\begin{definition}[\citet{matousek2013lectures}]\label{def:distortion} Fix any $\alpha \leq 1$.  A mapping $f: (\mathcal{Z}, \sigma) \to (\X, \rho)$ is called a \emph{$\frac{1}{\alpha}$-distortion embedding}, if there exist $c>0$ such that for all $u,v \in \mathcal{Z}$:
$$\alpha c \cdot \sigma(u,v) \leq \rho(f(u),f(v)) \leq c \cdot \sigma(u,v).$$
\end{definition}

It is not hard to see that $\alpha$-preservation is strictly weaker than $\frac{1}{\alpha}$-distortion.

\begin{observation}\label{obs:DistortionImpliesPreservationMetric}
    For any $\alpha \leq 1$. If $f$ constitutes an $\frac{1}{\alpha}$-distortion embedding\footnote{Note that $(\frac{1}{\alpha} < 1)$-distortion is impossible by definition.} of $ \mathcal{Z}$, then for all $R>0$, $f$ is an $(\alpha,R)$-preservation of $\mathcal{Z}$. In particular, if $\mathcal{Z}=G\in\Gn$ is equipped with the shortest path metric, then $f$ constitutes an $\alpha$-preservation of $G.$
\end{observation}

\begin{proof}\textbf{of Observation \ref{obs:DistortionImpliesPreservationMetric}}
    Let $f$ be a $(1/\alpha)$-distortion embedding. Then there exists $c>0$ such that\ for all $u,v \in \mathcal{Z}$: $c \cdot \sigma(u,v) \leq \rho(f(u),f(v)) \leq \frac{c}{\alpha} \cdot \sigma(u,v).$ Thus setting $r := cR/\alpha$:
    $$\sigma( u, v) < R \implies \rho(f_G(u), f_G(v)) < cR/\alpha = r$$
    $$\sigma( u, v) \geq R \implies \rho(f_G(u), f_G(v)) \geq cR = \alpha \cdot r. $$
\end{proof}
The alternate implication, $\alpha$-preservation implies $\frac{1}{\alpha}$-distortion, does not hold. In fact, the $\alpha$-preservation dimension and $\frac{1}{\alpha}$-distortion dimension can be arbitrarily far apart. For instance, for any $\alpha \leq 1,$ a $\frac{1}{\alpha}$-distortion embedding of a $n$-point unit simplex in $\mathbb{R}^n$ requires doubling dimension $\Omega\big(\frac{\log(n)}{\log(1/\alpha)}\big)$. On the other hand, the embedding that sends each point to a unique integer in $[n]$ is an $(\alpha,R)$-preservation for all $R>0$ (see proof of Proposition \ref{prop:alpha_gt_2}).



\section{Omitted Proofs}\label{app.proofs}

\subsection{Proofs from Preservation in General Metrics}
\label{app:proof_general_metrics}


\NeighborPreservingMain*

\begin{proof}\textbf{of Theorem \ref{thm:NeighborPreservingMain}(i)}
The proof proceeds by showing that $1-2^{-\Omega(n)}$ fraction of graphs in $\Gn$ have diameter at most $2$ and clique number at most $2\sqrt{n}$. Then, applying Observation \ref{prop:PGlowerbounds} and Lemma \ref{lem:preservation-hammer} yields the result.

\textit{Bounding $\gdiam(G)$.}  Corollary \ref{cor:gn_diam2} 
gives us
that with probability at least $1-n^2e^{-(n-1)/4}$ over a graph $G$ drawn uniformly from $\Gn$, we have that $\gdiam(G) \leq 2$.

\textit{Bounding $|\kappa(G)|$.} For any fixed $m \in [n]$ and $S \subseteq V$, define the random variables: (i) $X_m$ as the number of (not necessarily maximal) cliques of size $m$ in a graph $G$ drawn uniformly from $\Gn$ (identically, $G$ drawn from the Erdős–Rényi model over $n$ nodes, $\GnErdos$) and (ii) $Y_S = \indicate\big[G|_S \text{ is a clique}\big]$. Then by Markov's inequality:
\begin{align*}
\mathbb{P}_{G \sim \Unif(\Gn)}\Big[|\kappa(G)|&\geq m\Big] = \mathbb{P}_{G \sim \GnErdos}\Big[|\kappa(G)|\geq m\Big] = \mathbb{P}_{G \sim \GnErdos}\Big[X_m \geq 1\Big] \leq \mathbb{E}[X_m] \\   
& = \sum_{S\subseteq V: |S|=m} \mathbb{E}[Y_S] = \sum_{S\subseteq V: |S|=m} \left(\frac{1}{2}\right)^{\binom{m}{2}} = \binom{n}{m}\left(\frac{1}{2}\right)^{\binom{m}{2}}  \leq n^m 2^{-\binom{m}{2}}.
\end{align*}
Thus by picking $m=\big\lceil2\sqrt{n}\big\rceil$ gives that with probability at least $1-n^{(2\sqrt{n})}\cdot 2^{-\binom{2\sqrt{n}}{2}}$, we have that $|\kappa(G)| \leq 2\sqrt{n}$.

By combining these observations, we have that (when $n\geq 82$) for at least $1-2^{-n/5}$ fraction of graphs in $G\in \Gn$,
    $$\dim_\alpha(G) \geq \frac{\log(|P(G)|)}{\log(4\gdiam(G)/\alpha)} \geq \frac{\log(n/|\kappa(G)|)}{\log(8/\alpha)} \geq \frac{\log\left({\sqrt{n}}/{2}\right)}{\log(8/\alpha)} = \frac{\log\left(n\right) - 2\log(2)}{2\log(8/\alpha)}.$$
\end{proof}

\begin{proof}\textbf{Theorem \ref{thm:NeighborPreservingMain}(ii)} We will upper bound the diameter and the size of the largest clique of most graphs in $\Gnk$. Then, again applying Observation \ref{prop:PGlowerbounds} and Lemma \ref{lem:preservation-hammer} yields the result.

\textit{Bounding $\gdiam(G)$.}  Invoking Lemma \ref{lem:kregdiam} gives us
that with probability at least $1- O(n^{-\sqrt{k}+2})$ over a graph $G$ drawn uniformly from $\Gnk$, we have that $\gdiam(G) \leq \left\lceil \frac{\log (n-1)}{\log \left(\frac{k}{2\sqrt{k-1}+1/2}\right)}\right\rceil$. 

\textit{Bounding $|\kappa(G)|$.} Since no vertex has degree larger than $k$, all $G \in \Gnk$ has $|\kappa(G)| \leq k+1.$

By combining these observations, for any $n\geq 6$ and $k\geq 4$ even integers, we have that for at least $1-O(n^{-\sqrt{k}+2})$ fraction of graphs in $G\in \Gnk$,
$$\dim_\alpha(G) \geq \frac{\log(n/|\kappa(G)|)}{\log(4\gdiam(G)/\alpha)} \geq  \frac{\log(n/(k+1))}{\log\left(\frac{4}{\alpha}\left\lceil \frac{\log (n-1)}{\log \left(\frac{k}{2\sqrt{k-1}+1/2}\right)}\right\rceil\right)}
= \Omega\Bigg(\frac{\log(n/k)}{\log\log_k(n) + \log(4/\alpha)} \Bigg).$$
\end{proof}

\begin{proof}\textbf{of Proposition \ref{prop:nontrivial_PG_ub}}
For any \(G=(V,E)\in \Gn\), let $P(G)=\{P_1,\ldots,P_m\}$ be the $m$ parts of the clique partition. 

\emph{Case $\alpha < 1$.} 
Proposition \ref{prop:linf_ub} realizes $(\alpha<1)$-preservation in $\ell_\infty$ in at most $\Big\lceil\log_2(3)\Big\lceil \frac{ \log |P(G)|}{\log \lceil 1/\alpha\rceil}\Big\rceil \Big\rceil$ (doubling) dimensions.


\emph{Case $\alpha > 1$.}
Fix any $\epsilon>0$ small enough such that 
$\lceil\frac{1-\epsilon}{\alpha-1+\epsilon}\rceil = \lceil\frac{1}{\alpha-1}\rceil$. Consider an $n$-point space $\X$ (each point corresponding to a node in $V$), with distance $\rho$ defined as (for any $v,v'\in V$)
\[\rho(v,v') := \begin{cases}
        0 & v = v' \\
        \alpha & v\not\sim_G v'\\
        1- \epsilon & v\sim_G v',\  \{v,v'\}  \not\subseteq P_{i} \\
       \sigma_i(v,v') & \{v,v'\} \subseteq P_i  
    \end{cases}  \]
where $\sigma_i(v,v')$ is defined as follows. Let $\core(G|_{P_i}) = \{S_{i_1},\ldots,S_{i_k}\}$ be the $i_k$ parts of the neighborhood partition of $G|_{P_i}$. Recall that (see e.g.\ Observation \ref{obs:grid_packing}) $i_k$ points can be $(\alpha-1+\epsilon)$-packed 
in a $(1-\epsilon)$-diameter open ball $\ell_\infty^d$ (for $d =\lceil \log(i_k) / \log \lceil \frac{1-\epsilon}{\alpha-1+\epsilon}\rceil \rceil$). Thus mapping all vertices in $P_i$ to $i_k$ points such that all elements in the same neighborhood part $S_{i_j}$ are mapped to a single point and of different parts distance at least $(\alpha-1+\epsilon)$ apart (but contained within a $(1-\epsilon)$ diameter open ball), we can define $\sigma_i(v,v')$ as per the distances induced by this mapping (in $\ell_\infty$). It is instructive to note that either $\sigma_i(v,v')=0$ or $\alpha-1+\epsilon \leq \sigma_i(v,v') < 1-\epsilon$.

It is not hard to see that $\rho$ is a valid \emph{pseudo}-metric on $\X$. By standard metric identification one can make this into a bona fide metric that constitutes an $\alpha$-preservation of $G$. We will now determine this metric's doubling dimension. Let $K := \max_i i_k$ be the maximum number of neighborhood parts for any $P_i$ with $i\in[m]$, then:
\begin{itemize}
    \item Any ball of radius \(R < 1-\epsilon\) contains points only from one part $P_i$ for some $i\in [m]$. By construction it can be covered by $2^{\log_2(3)\lceil \log(K) / \log \lceil \frac{1-\epsilon}{\alpha-1+\epsilon}\rceil \rceil}$ balls of radius $R/2$. 

    \item Any ball of radius \(R \geq 1 - \epsilon\) can be covered by \(|P(G)|\) (open) balls of radius \(1 - \epsilon\), each containing points from exactly one part. Again by construction, each such ball can be covered by $2^{\log_2(3)\lceil \log(K) / \log \lceil \frac{1-\epsilon}{\alpha-1+\epsilon}\rceil \rceil}$ balls of radius $(1-\epsilon)/2$. Hence we can cover the $R$ radius ball by $|P(G)|\cdot 2^{\log_2(3)\lceil \log(K) / \log \lceil \frac{1-\epsilon}{\alpha-1+\epsilon}\rceil \rceil}$ balls of radius $R/2$.
\end{itemize}
Since $\epsilon>0$ is chosen small enough such that $\lceil\frac{1-\epsilon}{\alpha-1+\epsilon}\rceil = \lceil\frac{1}{\alpha-1}\rceil$, any ball in our constructed metric space can always be covered by at most $|P(G)|\cdot 2^{\log_2(3)\lceil \log(K) / \log \lceil \frac{1}{\alpha-1}\rceil \rceil}$ balls of half the radius. Thus, our construction has doubling dimension \(\big\lceil\log_2 |P(G)| + \log_2(3)\lceil \log(K) / \log \lceil \frac{1}{\alpha-1}\rceil \rceil\big\rceil\).

\emph{Case $\alpha=1$.} For any integer $m>0$,
by the construction for $\alpha>1$ case above, we know that for $\alpha = 1+\frac{1}{m}$ we can $\alpha$-preserve in a metric space $\X_m$ such that any ball is covered by at most $|P(G)|\cdot 2^{\log_2(3)\lceil \log(K) / \log m \rceil}$ balls of half the radius. Taking $m\rightarrow \infty$ we have
$\dim_{(\alpha=1)}(\X_\infty) \leq \lceil\log_2 (3|P(G)|)\rceil$.
\end{proof}

\begin{proof}\textbf{of Corollary \ref{cor:metric_ub_typical}}
Let \(\Delta(G) < C\) for some constant \(C\). Combining Lemmas \ref{lem:preservation-hammer} and \ref{lem:HAMMER2} tells us
\begin{align*}
    \dim_\alpha(G) &\geq (1/2)\Big(\max_{U }\frac{\log |P(G|_{U})|}{\log(4\gdiam(G|_{U})/\alpha)} +  \indicate[{\alpha > 1}]\cdot \max_{U}\;\frac{\log \big| \core(G|_U) \big|}{\log(\frac{4\gdiam(G|_U)}{\alpha - 1})}\Big) \\
    &\geq (1/2)\Big(\max_{U}\frac{\log |P(G|_{U})|}{\log(4C/\alpha)} +  \indicate[{\alpha > 1}]\cdot\max_{U}\;\frac{\log \big| \core(G|_U) \big|}{\log(\frac{4C}{\alpha - 1})}\Big) \\
    &= (1/2)\Big(\frac{\log |P(G)|}{\log(4C/\alpha)} +  \indicate[{\alpha > 1}]\cdot\frac{\log \big| \core(G) \big|}{\log(\frac{4C}{\alpha - 1})}\Big),
\end{align*}
where the last line follows from monotonicity of clique and neighborhood partitions; for \(U\subseteq W\), \({|P(G|_U)| \leq |P(G|_W)|}\) and \({|C(G|_U)| \leq |C(G|_W)|}\). Meanwhile, Proposition \ref{prop:nontrivial_PG_ub} tells us
   \[\dim_\alpha(G) = O\Big(\frac{\log |P(G)|}{\log(4/\alpha)} + \indicate[{\alpha > 1}]\cdot  \frac{\max_{S\in P(G)} \log |C(G|_S)|}{\log(\frac{8}{\alpha - 1})}\Big).\]
  Monotonicity tells us \(\max_{S\in P(G)} \log |C(G|_S)| \leq \log |C(G)|\), completing the proof.

\end{proof}

\newpage

\subsection{Proofs from Preservation in Normed Spaces}
\label{app:proof_normed_spaces}

\polyMain*






\begin{proof}
\textbf{of Theorem \ref{thm:polyMain}(i)}
Consider the set $S \subset \Gn$ of all graphs with diameter at most $2$. If $n\geq 82$, we have (see Corollary \ref{cor:gn_diam2}) 
$|S| \geq (1-n^2 e^{-(n-1)/4})|\Gn|\geq  (1-2^{-n/5})|\Gn| \geq 2^{n \choose 2}/2$.
A straightforward application of Lemma \ref{lem:family_lb} on $S$ yields that there exists at least one graph that must require at least $\frac{({n\choose 2}-1) \log(2)}{n\log(16/(\alpha - 1))} = \Omega(n/\log(\frac{16}{\alpha-1}))$ (doubling) dimensions to $(\alpha>1)$-preserve any normed space. But we can do better. 

Define the set of ``low-embedding'' graphs:
$$T := \Big\{G\in \Gn \;\big| \dim_\alpha(G,\mathbb{L}) \leq \frac{(n/3)\log(2)}{\log(16/(\alpha-1))} 
 =:d\Big\}.$$
 
Assume towards contradiction that $|T| \geq 2^{-n/6} |\Gn|$.
%
Then observe that 
$|T\cap S| = |T\setminus S^c| \geq |T| - |S^c| \geq (2^{-n/6} - 2^{-n/5})|\Gn| \geq 2^{-n/5}|\Gn| = 2^{{n\choose 2} - (n/5)}$ (when $n\geq 82$). Thus $\dim_\alpha(T\cap S,\mathbb{L}) \leq \dim_\alpha(T,\mathbb{L})\leq d$ (see Proposition \ref{prop:monotonicity}), but by Lemma \ref{lem:family_lb} $\dim_\alpha(T\cap S,\mathbb{L}) \geq \frac{({n\choose 2}-\frac{n}{5}) \log(2)}{n\log(16/(\alpha - 1))} > d$ (for $n\geq 5$). This implies there exists a graph \(G\in T\cap S \subseteq T\) where $\dim_\alpha(G) > d$, arriving at a contradiction.

Therefore\footnote{The constants in the proof can be improved.} (when $n\geq 82$), at least $(1-2^{-n/6})$ fraction of graphs in $\Gn$ require at least $d =\frac{(n/3)\log(2)}{\log(16/(\alpha-1))} $ (doubling) dimensions to be $(\alpha>1)$-preserved in any normed space.
\end{proof}

\begin{restatable}{observation}{VectorCentered} \label{lem:vector_centered}
    Let \(S\subseteq \Gn\). Suppose \(f:S\to\X^n\) \(\alpha\)-preserves \(S\) in a normed space \(\X\). Then there exists \(f': S\to\X^n\) which \(\alpha\)-preserves \(S\) such that for all $G\in S$
\begin{itemize}
    \item (equal scaling) The neighborhood threshold for $f'(G)$ is exactly $1$ (cf.\ Definition \ref{def:preserve}).
    \item (centeredness) \((1/n)\sum_{v \in V} f_G'(v) = \vec 0\). 
\end{itemize}
\end{restatable}

\begin{restatable}{lemma}{FamilyLB} \label{lem:family_lb}
    Let $\mathbb{L}$ be the collection of all normed spaces. For any set \(S\subseteq \Gn\) of graphs on $n$ vertices with diameter at most \(R\), and any \(\alpha \in (1,2)\),
    $$\dim_{\alpha}(S, \mathbb{L})  \geq \frac{\log |S|}{n \log(8R/(\alpha-1)) }.$$
\end{restatable}
\begin{proof}\textbf{of Lemma \ref{lem:family_lb}} WLOG assume $R<\infty$, and $f: S \rightarrow \X^n$ be an $(\alpha>1)$-preserving map into a normed space $(\X,\|\cdot\|)$ such that for all $G\in S$, $f(G)$ produces an embedded centered at the origin and has the neighborhood threshold of $1$ (cf.\ Observation \ref{lem:vector_centered}).
Let \(B_R(\vec {0}) \subset \X\) be the open ball of radius \(R\) about the origin. Observe that \(\{f_G(v): G\in S, v\in V\} \subset B_R(\vec {0})\), since for all \(G\in S\) and \(v\in V\),
\[\|f_G(v)\| 
= \bigg\|f_G(v) - \frac{1}{n} \sum_{u\in V} f_G(u) \bigg\| \leq (1/n) \sum_{u\in V} \|f_G(v) - f_G(u)\| \leq \diam(f(G)) < 1\cdot \gdiam(G)\leq R.\]
Fix an \(((\alpha - 1)/4)\)-cover \(N\) of \(B_R(\vec {0})\). Per Observation \ref{obs:covering_number}, we may assume $|N| \leq (8R/(\alpha-1))^{\dim(\X)}$. Define the projection map \(\Phi: \mathcal{X}^n \to N^n\) by \(\Phi((x_i)_{i\in [n]}) = (\argmin_{p\in N} \|p-x_i\|)_{i\in [n]}\), where ties are broken in an arbitrary but consistent way. 
    
  Consider the map from graphs to their projections onto the net, \(\Phi \circ f: S \to N^n\). We claim this map is injective. Suppose, for sake of contradiction, it was not injective. Then there exists distinct \(G, G' \in S\) such that \(\Phi(f(G)) = \Phi(f(G'))\). Let \((u,v)\) be a pair on which $G$ and $G'$ disagree on an edge. Without loss of generality, we may assume $u\nsim_G v$ and $u\sim_{G'} v$, thus
\[\|f_G(u) - f_G(v)\| \geq \alpha \textup{ \quad \quad  and \quad \quad } \|f_{G'}(u) - f_{G'}(v)\| < 1.\]

For convenience, let \(\Phi(p) \in  N\) denote the projection of a point \(p\) onto the net. Then by assumption we have \(\Phi(f_G(u)) = \Phi(f_{G'}(u)) =: N_u\) and \(\Phi(f_G(v)) = \Phi(f_{G'}(v)) =: N_v\). Then we have
\begin{align*}
   \|f_G(u) - f_G(v)\| &\leq 
\|f_G(u) - N_u \| + \|N_u - f_{G'}(u)\| \\
&+ \|f_{G'}(u) - f_{G'}(v)\| + \|f_{G'}(v)- N_v\| + \|N_v - f_G(v)\| \\ 
&< \frac{\alpha - 1}{4} + \frac{\alpha - 1}{4} + 1 + \frac{\alpha - 1}{4} + \frac{\alpha - 1}{4} = \alpha.
\end{align*}
This contradicts the fact that \(\|f_G(u)- f_G(v)\| \geq \alpha\) and thereby establishes injectivity. Therefore we have
\begin{align*}
   |S| &\leq |N^n|  = |N|^n = \Big(\frac{8R}{\alpha - 1}\Big)^{n \dim(\X) } . 
\end{align*}
Rearranging the expression, we have that, for any \(f\) which \((\alpha>1)\)-preserves \(S\) in a normed space \(\X\), the doubling dimension of \(\X\) is at least \(\frac{\log |S|}{n \log(8R/(\alpha-1)) }\).
\end{proof}



\begin{proof}\textbf{of Theorem \ref{thm:polyMain}(ii)} 
A theorem in Section 3.1 of \citep{reiterman1989embeddings} says that for \(n\geq 38\),  (\(1-\frac{1}{n}\))-fraction of \(G\in \Gn\) have sphericity at least \(n/15-1\). This lower bound carries over for for \((\alpha=1)\)-preservation in \(\ell_2\), since \(\dim(\ell_2^d)\leq d\).  We strengthen this result by showing the lower bound applies to \((1-\frac{1}{2^n})\)-fraction of \(\Gn\). We use a similar argument which relies on upper bounds for consistent sign assignments of bounded-degree polynomials. 

    We specialize Lemma \ref{lem:Poly-Hammer} to the $p=2$ case. Fix any $S \subseteq \Gn$ with $S \geq 2^{-n}|\Gn|$ such that some $f:\Gn \to \ell_2^d$  $(\alpha=1)$-preserves $S$. Then by Lemma \ref{lem:Poly-Hammer}:
    \begin{align*}
       && |S| &\leq \left(\frac{4e(n-1)}{d}\right)^{nd} \\
        &\implies &2^{\binom{n}{2}-n} = 2^{-n} |\Gn| &\leq \left(\frac{4e(n-1)}{d}\right)^{nd} \\
        &\iff &\left(\frac{n-3}{2}\right)\log 2 &\leq d \log \left(\frac{4e(n-1)}{d}\right) 
        = d \left( \log \left(\frac{e(n-1)\log 2}{10d}\right) + \log \left(\frac{40}{\log 2}\right) \right)\\
        &\implies & \left(\frac{n-3}{2}\right)\log 2 &\leq \left(\frac{n-1}{10}\right)\log 2 + d\log \left(\frac{40}{\log 2}\right)  \\
        &\iff & d &\geq \frac{2n-7}{5\log_2 \left(\frac{40}{\log 2}\right)} \geq \frac{n}{15}-\frac{1}{4}.
    \end{align*}

    Thus no more than $2^{-n}$ fraction of $\Gn$ can be $(\alpha=1)$-preserved in $\ell^d_2$ for $d < \frac{n}{15}-\frac{1}{4}.$ To finish the proof, it is noted that the doubling dimension of $\ell_2^d$ is at least $d$.
\end{proof}

\begin{lemma}\label{lem:Poly-Hammer}
    For  $p\geq2$ even integer and $S \subseteq \Gn$, if  there exists $f: \Gn \to \ell_p^d$ that $(\alpha=1)$-preserves $S$, then:
    $$|S| \leq \left(\frac{2ep(n-1)}{d}\right)^{nd}.$$
\end{lemma}



\begin{proof} \textbf{of Lemma \ref{lem:Poly-Hammer}}
    For any $G = (V,E) \in S$. Since $f$ $(\alpha=1)$-preserves $G$, there exists $r>0$ and $f_G:V \to \mathbb{R}^d$ such that:
    
    $$(u,v) \in E \implies \lVert f_G(u)-f_G(v)\rVert_p < r$$
    $$(u,v) \not\in E \implies \lVert f_G(u)-f_G(v)\rVert_p \geq r.$$

Pick $\epsilon_n>0$ such that for all $G =(V,E)\in S$ and $(u,v)\in E$, $\lVert f_{G}(u)-f_{G}(v)\rVert_q<r-\epsilon_n$ which exists since $|S|$ and $n<\infty$ are finite. Then,

    $$\{P_{uv}(x,x') = \lVert x - x'\rVert^p_p -r+\epsilon_n\ |\ x,x'\in\mathbb{R}^d,\ u,v \in V, u\neq v\}$$

    is a set of $\binom{n}{2}$ polynomials (since $p$ is even) over $nd$ variables with the property that for all $u,v \in V, u\neq v$:

    \begin{itemize}
        \item[(i)] If $(u,v)\in E$, then $P_{uv}(f_G(u),f_G(v)) <0$.
        \item[(ii)] If $(u,v)\not\in E$, then $P_{uv}(f_G(u),f_G(v)) >0$.
    \end{itemize}

    Therefore, each $G \in S$ will produce a unique non-zero sign assignment of $\{P_{uv} |u,v \in V, u\neq v\}$, so by Lemma \ref{lem:warrens}:

    $$|S| \leq \left( \frac{4ep\binom{n}{2}}{nd}\right)^{nd} = \left( \frac{2ep(n-1)}{d}\right)^{nd}.$$
\end{proof}



\begin{lemma}
[\cite{WarrensPoly}] 
\label{lem:warrens}
If \(\{p_1,\ldots,p_m\}\) is a set of polynomials of degree at most \(D\geq 1\) in \(N\) variables with \(m\geq N\), then the number of consistent non-zero sign assignments to the \(p_i\) is at most \((4eDm/N)^N\), that is, $\Big|\Big\{ \big(\textup{sign}(p_1(x)), \dots, \textup{sign}(p_m(x))\big) \;:\; x \in \mathbb{R}^N \text{ s.t. } p_i(x)\neq 0 \text{ for } i\in[m] \Big\}\Big| \leq \big(4eDm/N\big)^N$.
\end{lemma}


\begin{proof}\textbf{of Proposition \ref{prop:chiGBasedUpperBound}}
    Fix any $G\in \Gn$, and let $P(G)=\{P_1,\ldots,P_m\}$ be the $m$ parts of its clique partition. WLOG, assume each vertex in \(G\) has degree at least \(1\) (since isolated vertices can be embedded separately).

\emph{Case $\alpha\in(0,\frac{1}{\sqrt{2}})$.}
A simple $\alpha$-packing of $m$ points in a unit $\ell_2$ ball achieves the desired $\alpha$-preserving embedding. Specifically, by Observation \ref{obs:grid_packing}(ii) (take $n=m$, $r=1$ and $\epsilon = \alpha$) we know that $m$ points can be $\alpha$-packed in an (open) unit ball in $\ell_2^d$ with $d = \lceil \frac{4 \log (m+1)}{2-4\alpha^2}\rceil$. Thus, the mapping where all vertices of the partition $P_i$ of the input graph $G$ is mapped to the $i$-th point of the packing in unit ball yields an $\alpha$-preserving embedding in $\ell_2$, with (doubling) dimension at most $\Big\lceil\log_2(5) \Big\lceil \frac{ 4 \log (|P(G)|+1)}{2-4\alpha^2}\Big\rceil \Big\rceil$.
\\

\emph{Case $\alpha\in(\frac{1}{\sqrt{3}},1)$.} A low-distortion embedding of a regular simplex of $m$ points in $\ell_2$ can be used to get a good $(\alpha<1)$-preservation. In particular, consider a regular unit simplex of $m$ points in $\ell_2$. 
By Lemma \ref{lem:JL} we know that for any $0<\epsilon\leq \frac{1}{2}$, a $\sqrt{\frac{1+\epsilon}{1-\epsilon}}$-distortion embedding of these $m$ points exists in $\ell_2^d$ with $d = \lceil\frac{12}{\epsilon^2}\log m\rceil$. Hence by picking $\epsilon = \frac{1-\alpha^2}{1+\alpha^2}$, we get an $\alpha$-distortion and therefore an $\alpha$-preservation in $\ell_2^d$ for $\alpha\in(\frac{1}{\sqrt{3}},1)$, cf.\ Observation \ref{obs:DistortionImpliesPreservationMetric}.
\\

\emph{Case $\alpha\in\Big[1,\frac{1}{\sqrt{1-\min(1,({1}/{4\lambda_G}))}}\Big)$.} Fix an arbitrary $G \in \Gn$. 
Let $A = A(G)$ be the adjacency matrix of $G$ with maximum eigenvalue $\lambda_G$, and let $A^c$ be the adjacency matrix of the complement graph. Define ${D := A^c + (1-\frac{1}{\lambda_G})A}$ as a squared interpoint distance matrix over the $n$ vertices. Using a theorem of Schoenberg \citep{bavaud2011schoenberg}, we can verify that the (non-squared) distances of $D$ are $\ell_2^n$ isometrically embeddable: for $u\in \mathbb{R}^n$ with $\|u\|_2 = 1$ and $\textbf{1}^Tu = 0$, we have
    $$u^TDu = u^T\Big(\textbf{1}\textbf{1}^T - I_n - \frac{1}{\lambda_G}A\Big)u = - \lVert u \rVert_2^2 - \frac{1}{\lambda_G}u^TAu \leq - 1 + \frac{1}{\lambda_G} |u^TAu| \leq 0.$$
Hence the same $\ell_2^n$ embedding is also an \(\big(\alpha' = (1-1/\lambda_G)^{-1/2}\big)\)-preservation of \(G\): if \(i\not\sim j\), then \(D_{ij} = 1\), whereas if \(i\sim j\), then \(D_{ij} = (1-1/\lambda_G)^{-1/2}\). Given this realization of points in \(\mathbb{R}^n\), we can apply Lemma \ref{lem:JL} (with \(\epsilon = \frac{1}{4\lambda_G} \leq \frac{1}{4}\); note that \(\lambda_G\) is at least the average vertex degree, which is at least \(1\) in this graph by assumption) to conclude neighborhood preservation of $G$ is possible in \(\mathbb{R}^d\) with $d = \lceil 12\cdot 16\lambda_G^2 \log n\rceil$ with a slight degradation in the $\alpha$-parameter. In particular, $\alpha$-preservation is possible in $\mathbb{R}^d$ for $\alpha\in\Big(1, \big(1-\frac{1}{4\lambda_G}\big)^{-1/2}\Big) \subseteq \Big(1,\alpha'\sqrt\frac{1-\epsilon}{1+\epsilon}\Big)$, cf.\ Observation \ref{obs:DistortionImpliesPreservationMetric}. The extra factor of \(\log_2(5)\) is an artifact of switching from \(\ell_2\) dimension to doubling dimension. The proposition statement follows by replacing \(G\) with \(G/C(G)\) (see proof of Proposition \ref{prop:linf_ub} for the definition), therefore \(n\) with \(|C(G)|\), in the argument.
\\

\emph{Case $\alpha > (1- \frac{1}{\lfloor n/2\rfloor})^{-1/2}$}. For simplicity of our discussion assume $n$ is even, and consider a complete bipartite graph \(G \in \Gn\) with parts $S_0$ and $S_1$ each containing $n/2$ vertices. For convenience label the vertices $1,\ldots, n$ (in any order). 

We shall show that if an $\alpha$-preservation of $G$ exists in $\ell_2$ then necessarily $\alpha \leq (1-\frac{1}{\lfloor n/2\rfloor})^{-1/2}$.
Let \(\{x_1,\ldots,x_n\}\) be an \(\alpha\)-preservation of this graph in $\ell_2$ with neighborhood threshold $r$ (see Definition \ref{def:preserve}). WLOG we can assume the embedding resides in \(\mathbb{R}^n\).
Let \(u \in \mathbb{R}^n\) be such that \(u_i = \textbf{1}[i\in S_0] - \textbf{1}[i\in S_1]\). Clearly \(u^T \textbf{1} = 0\). Then by a theorem of Schoenberg \citep{bavaud2011schoenberg}, \(u^T Du \leq 0\), where \(D\in \mathbb{R}^{n\times n}\) is the squared interpoint distance matrix: \(D_{ij} = \|x_i-x_j\|_2^2\).
By definition of \(u\), this implies $ \sum_{i\not\sim j} D_{ij} \leq \sum_{i\sim j} D_{ij}$. Applying the definition of \(\alpha\)-preservation (i.e.\ embedded pairwise distance of any edge-connected pair of vertices is at most $r$, otherwise is at least $\alpha r$)
, and counting the number of edges and non-edges we have
\[ \alpha^2r^2(n^2/2 - n) \leq \sum_{i\not\sim j} D_{ij} \leq \sum_{i\sim j} D_{ij} \leq  r^2 n^2/2. \]
The requirement on $\alpha$ for $\ell_2$ preservation follows.
When \(n \geq 4\) is odd, a similar analysis holds: let \(G\) be a complete bipartite graph on $n-1$ (even) vertices and let the remaining \(n\)-th vertex have no edges. Define \(u\) the same as before with \(u_n = 0\). The same analysis yields the stated restriction on $\alpha$.
%
\end{proof}

\subsection{Proofs from Preservation of Clustered Data}
\label{app:cluster}

\plantedPartition*

\begin{proof}\textbf{of Theorem \ref{thm:planted_partition}} The proof of the theorem follows from showing (for any $G=(V,E) \sim  \PP_{p,q}(S_1,\ldots,S_k)$):  (i) $\gdiam(G) \leq 2$ with probability at least $1 - \exp(-\Omega(n))$, and (ii) $|\kappa(G)| \leq \left( \frac{2cn}{k}\right)^{\xi_{p,q}}$ with probability at least ${1 - \exp(- \Omega(n^{2p+2q-2pq}))}$.
Thus by applying Lemma \ref{lem:preservation-hammer} and Proposition \ref{prop:PGlowerbounds} we get that for $G \sim \PP_{p,q}(S_1,\ldots,S_k)$:
$$\mathbb{P}\left[\dim_\alpha(G) < \frac{1}{\log\left(8/\alpha\right)}\log\left(\frac{n}{(2cn/k)^{\xi_{p,q}}}\right)\right] \leq \mathbb{P}\left[\text{$G$ has $|\kappa(G)| > \left( \frac{2cn}{k}\right)^{\xi_{p,q}}$ or $\gdiam(G) > 2$}\right]$$
$$\leq \exp(- \Omega(n^{2p+2q-2pq})) + \exp(-\Omega(n)) \leq \exp(- \Omega(n^{\min(2p+2q-2pq,1)})).$$

\emph{Bounding the diameter.} The diameter bound follows directly from Lemma \ref{lem:diam2graphs}: (since $0 \leq q\leq p$) with probability at least \(1 - n^2 e^{-q^2(n-1)}\), the diameter of \(G\) is at most \(2\). 
\\

\emph{Bounding the clique number.} We bound the probability that $G \sim  \PP_{p,q}(S_1,\ldots,S_k)$ contains a clique of size greater than $m:= \lceil \left( \frac{2cn}{k}\right)^{p+q-pq} \rceil $. For any $S\subseteq V$ and $G=(V,E) \sim \PP_{p,q}(S_1,\dots,S_k)$, define the random variables ${Y_S := \indicate[G|_S \text{ is a clique}]}$. Then $C_m := \sum_{S \subseteq V : |S|=m} Y_S$ is the number of $m$-sized cliques in $G$. Therefore:

\begin{align*}
\mathbb{P}_{G \sim \PP_{p,q}(S_1,\ldots,S_k)}\Big[\text{$G$ contains an $m$ }&\text{sized clique}\Big] = \mathbb{P}[C_m \geq 1 ] 
\leq \E[C_m] = \sum_{S \subseteq V : |S|=m} \E[Y_S] \\
&= \sum_{S \subseteq V : |S|=m} \mathbb{P}(G|_S \text{ is a clique}) \\
&=  \sum_{S \subseteq V : |S|=m} p^{\binom{|S\cap S_1|}{2}+ \cdots + \binom{|S\cap S_k|}{2}}q^{\sum_{1\leq i<j\leq k}|S\cap S_i|\cdot |S\cap S_j|}\\
&= q^{\binom{m}{2}}\sum_{S \subseteq V : |S|=m} (p/q)^{\binom{|S\cap S_1|}{2}+ \cdots + \binom{|S\cap S_k|}{2}}, 
\end{align*}
where \({\sum_i x_i \choose 2} = 
\sum_i {x_i\choose 2} + \sum_{i < j} x_i x_j\) holds for all \(x_1,\dots,x_n\) non-negative integers. Define:
$$A := \{S \subseteq V: |S|=m, \exists i \in[k] \text{ s.t.\ } |S\cap S_i| > m/2\},
$$  
$$B := \{S \subseteq V: |S|=m, \forall i \in[k] \text{ s.t.\ } |S\cap S_i| \leq m/2\}.$$
Note that $A\sqcup B = \{S \subseteq V: |S|=m\}$. Continuing onwards, we have
\begin{align*}
    &= q^{\binom{m}{2}}\left(\sum_{S \in A} (p/q)^{\binom{|S\cap S_1|}{2}+ \cdots + \binom{|S\cap S_k|}{2}}+\sum_{S \in B} (p/q)^{\binom{|S\cap S_1|}{2}+ \cdots + \binom{|S\cap S_k|}{2}}\right) \\
    &\leq q^{\binom{m}{2}}\left(|A|(p/q)^{\binom{m}{2}} + |B|(p/q)^{(\frac{m^2}{4}-\frac{m}{2})}\right), 
\end{align*}
where for the first exponent, we use \(\sum_i {x_i \choose 2} \leq {\sum_i x_i \choose 2}\), and for the second exponent, we use Hölder's inequality: \(\langle x,y \rangle \leq \|x\|_\infty\|y\|_1\), therefore \(\sum_{i} \binom{|S\cap S_i|}{2} = \frac{1}{2}\sum_{i}( |S\cap S_i|-1)|S\cap S_i| \leq \frac{1}{2}\left(\frac{m}{2}-1 \right)m
\). We can upper bound \(|B|\) by \({n\choose m}\) and \(|A|\) by \({n\choose m}\cdot \textbf{1}[\frac{m}{2} < \frac{cn}{k}]\) because \(\max_i |S\cap S_i| \leq  \max_i |S_i| \leq cn/k\). Thus,
\begin{align*}
    &\leq q^{\binom{m}{2}}\Big({n\choose m}(p/q)^{\binom{m}{2}}\cdot \indicate\Big[m < \frac{2cn}{k}\Big] + \binom{n}{m}(p/q)^{(\frac{m^2}{4}- \frac{m}{2})} \Big) \\
      &= {n\choose m} p^{m\choose 2}\Big( \indicate\Big[m< \frac{2cn}{k}\Big] +  (q/p)^{m^2/4}\Big).
\end{align*}

From here we can evaluate two cases:
\begin{itemize}
    \item If \(m = \lceil \left( \frac{2cn}{k}\right)^{p+q-pq} \rceil  \geq 2cn/k\), then our bound is \({n\choose m} q^{m^2/4}p^{m^2/4-m/2} = \exp(-\Omega(n^{2p+2q-2pq}))\). 
    \item If \(m = \lceil \left( \frac{2cn}{k}\right)^{p+q-pq} \rceil  < 2cn/k\), then it must be the case that \(p<1\), so our upper bound becomes \({{n\choose m} \Big(p^{m\choose 2} + p^{(\frac{m^2}{4} - \frac{m}{2})}
    q^{m^2/4}\Big) = \exp(-\Omega(n^{2p+2q-2pq}))}\).
\end{itemize}
Therefore, $\mathbb{P}_{G \sim \PP_{p,q}(S_1,\ldots,S_k)}\Big[ |\kappa(G)|\geq m\Big] \leq \exp(- \Omega(n^{2p+2q-2pq}))$.
\end{proof}


\begin{proof}\textbf{of Proposition \ref{prop:clusters_alpha_geq_1}}
    For any $i\in [k]$ and $u,v \in S_i$, the probability that $N(u) = N(v)$ (see Definition \ref{def:CG}) is at most
    $$ (p^2+(1-p)^2)^{|S_i|} \cdot (q^2+(1-q)^2)^{n-|S_i|} \leq (q^2+(1-q)^2)^{n(1-c/k)} \leq \max(q,1-q)^{2n(1-c/k)}.$$
    For any $u \in S_i$ and $v \in S_j$ with $i \neq j$, the probability that $N(u)=N(v)$ is at most
    $$ (pq+(1-p)(1-q))^{|S_i|+|S_j|} \cdot (q^2+(1-q)^2)^{n-|S_i|-|S_j|} \leq \max(q,1-q)^{2n}. $$

    Applying Lemma \ref{lem:diam2graphs}, with probability at least $1 - n^2\left(\max(q,1-q)^{2n(1-c/k)}+e^{-q^2(n-1)}\right)$,
    we have ${|C(G)| = n}$ and $\gdiam(G)\leq 2$. The lower bound follows from applying Lemma \ref{lem:HAMMER2}.
\end{proof}



\end{document}